\documentclass[11pt]{article}

\usepackage[final]{acl}

\usepackage{times}
\usepackage{latexsym}

\usepackage[T1]{fontenc}

\usepackage[utf8]{inputenc}

\usepackage{microtype}

\usepackage{inconsolata}

\usepackage{graphicx}

\usepackage{amsmath}
 \usepackage{amssymb}
\usepackage{booktabs}
\usepackage{tcolorbox}
\usepackage{multirow}
\usepackage[normalem]{ulem}
\useunder{\uline}{\ul}{}
\usepackage{subcaption}
\usepackage{enumitem}
\usepackage{graphicx}
\usepackage{listings}
\usepackage{xcolor}
\usepackage{colortbl}
\usepackage{array}
\usepackage{hyperref}
\usepackage{authblk} 
\usepackage[capitalize,noabbrev]{cleveref}
\tcbuselibrary{listings,breakable,skins}

\usepackage{fvextra}

\crefname{section}{Sec.}{Secs.}
\Crefname{section}{Sec.}{Secs.}

\crefname{appendixsec}{Appendix}{Appendices}
\Crefname{appendixsec}{Appendix}{Appendices}

\setlist{nosep}

\lstdefinelanguage{json}{
  morestring=[b]",
  morecomment=[l]{//},
  morecomment=[s]{/*}{*/},
  showstringspaces=false,
  breaklines=true,
  basicstyle=\footnotesize\ttfamily,
}
\tcbset{jsonstyle/.style={
  enhanced,
  breakable,
  colback=black!2,
  colframe=black!55,
  boxrule=0.4pt,
  arc=1.2pt,
  left=6pt,
  right=6pt,
  top=6pt,
  bottom=6pt,
  fonttitle=\small,
  coltitle=black,
  listing engine=listings,
  listing only,
  listing options={language=json},
}}

\title{Multi-Field Tool Retrieval}

\author[1,2]{\textbf{Yichen Tang}\thanks{tangyc21@mails.tsinghua.edu.cn}}
\author[1]{\textbf{Weihang Su}}
\author[1]{\textbf{Min Zhang}}
\author[1]{\textbf{Yiqun Liu}}
\author[2,1]{\textbf{Qingyao Ai}\thanks{Corresponding Author: aiqy@tsinghua.edu.cn}}

\affil[1]{Department of Computer Science and Technology, Tsinghua University}
\affil[2]{Quancheng Laboratory}

\begin{document}
\maketitle
\begin{abstract}
  Integrating external tools enables Large Language Models (LLMs) to interact with real-world environments and solve complex tasks.
  Given the growing scale of available tools, 
  effective tool retrieval is essential to mitigate constraints of LLMs' context windows and ensure computational efficiency. 
  Existing approaches typically treat tool retrieval as a traditional ad-hoc retrieval task, matching user queries against the entire raw tool documentation.
  In this paper, we identify three fundamental challenges that limit the effectiveness of this paradigm: (i) the incompleteness and structural inconsistency of tool documentation; (ii) the significant semantic and granular mismatch between user queries and technical tool documents; and, most importantly, (iii) the multi-aspect nature of tool utility, that involves distinct dimensions, such as functionality, input constraints, and output formats, varying in format and importance.
  To address these challenges, we introduce Multi-Field Tool Retrieval, a framework designed to align user intent with tool representations through fine-grained, multi-field modeling. 
  Experimental results show that our framework achieves SOTA performance on five datasets and a mixed benchmark, exhibiting superior generalizability and robustness~\footnote{We have open-sourced all the code and prompts in \url{https://github.com/LittleDinoC/MFTR}.}.
\end{abstract}

\section{Introduction}
\label{sec:introduction}

Large language models (LLMs) have demonstrated remarkable capabilities in natural language tasks~\cite{LLMareFewshotLearners, JMLR:v24:22-1144, 10.1145/3744746}, yet standalone deployment faces critical challenges: the lack of grounding to interact with real-world environments~\cite{bisk-etal-2020-experience, Ahn2022DoAI}, and the lack of efficient capabilities for continuous learning and long-term accumulation of functional skills~\cite{10.1145/3735633, Wang2023VoyagerAO}.
A natural solution is to extend LLMs with external tools~\cite{toolformer, Qin2023ToolLW}. 
Tool augmentation enables LLMs to actively interact with external environments~\cite{Mialon2023AugmentedLM} while managing capabilities outside model parameters~\cite{toolbench, apigen}. 
With recent advances in reinforcement learning and agents, enabling LLMs to use a few specific tools is no longer difficult~\cite{liu2025toolace, zeng2025toolacermodelawareiterativetraining, yao2023react, anthropic2024mcp}. 
Thus, efficient tool management has become an increasingly important bottleneck for general-purpose agents.

As of today, one of the most popular tool management methods is tool retrieval.
Modern tool repositories are highly heterogeneous and dynamic, comprising manually designed APIs, transformed web services, and even LLM-generated tools ~\cite{Wang2023VoyagerAO, gorilla, tang2023toolalpacageneralizedtoollearning}.
Constrained by LLMs' limited context windows and computational complexity, tool retrieval aims to select only the most relevant tools given the task context.
Most existing methods formulate tool retrieval as an ad-hoc retrieval problem: representing each tool using its documentation and retrieving tools according to user queries. 
Previous studies mainly focus on retrieval workflows~\cite{liu2024toolnetconnectinglargelanguage, pluto}, representation optimization~\cite{chen-etal-2024-invoke, onlinerag, yuan-etal-2025-easytool}, and specialized retrievers~\cite{toolbench, colt}.

\begin{figure*}[t]
    \centering
    \begin{subfigure}[b]{0.48\textwidth}
        \centering
        \includegraphics[width=\linewidth]{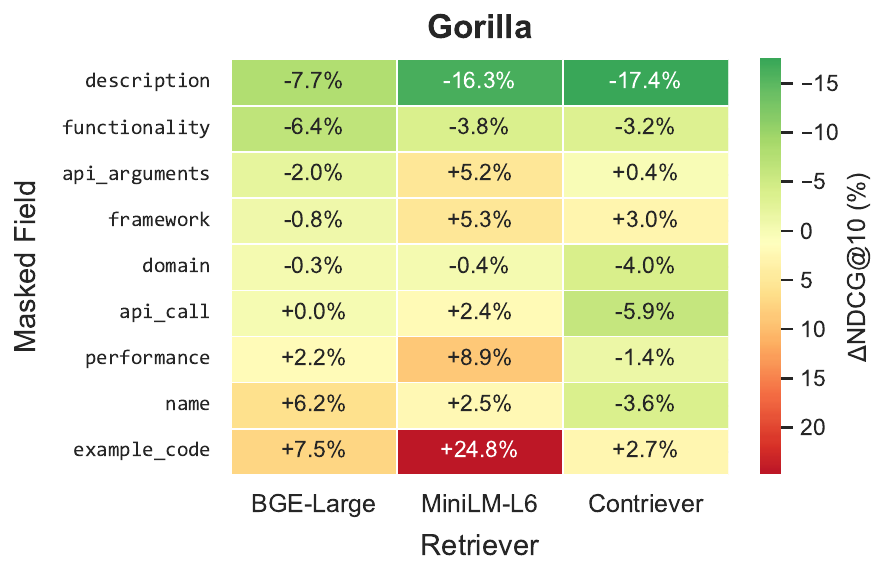}
        \label{fig:mask_field_gorilla}
    \end{subfigure}
    \hfill
    \begin{minipage}[b]{0.48\textwidth}
        \centering
        \begin{subfigure}[b]{\textwidth}
            \centering
            \includegraphics[width=\linewidth]{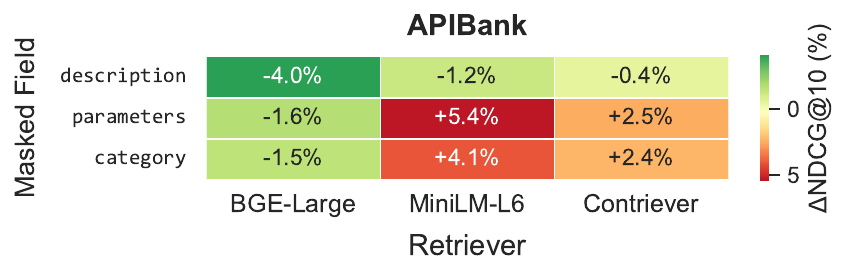}
            \label{fig:mask_field_apibank}
        \end{subfigure}

        \vspace{-5.5mm} 

        \begin{subfigure}[b]{\textwidth}
            \centering
            \includegraphics[width=\linewidth]{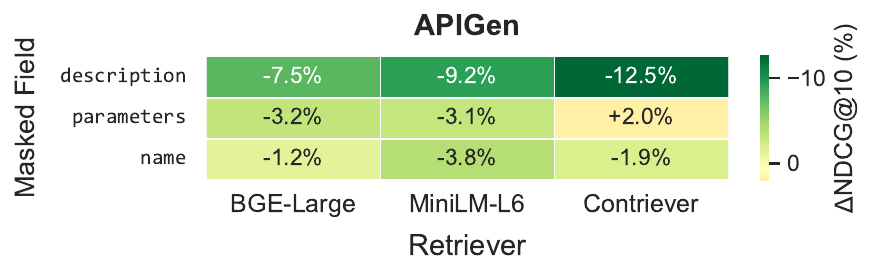}
            \label{fig:mask_field_apigen}
        \end{subfigure}
    \end{minipage}
    
    \vspace{-4mm}

    \caption{
    Impact of field masking on retrieval performance across Gorilla~\cite{gorilla}, APIBank~\cite{apibank}, and APIGen~\cite{apigen} datasets.
    The three datasets have incompatible tool documentation formats.
    The results reveal significant differences in field contributions and retriever preferences, indicating that treating tool documentation as a whole unit is suboptimal.
    }
    \label{fig:mask_field_experiment}

    \vspace{-2mm}
\end{figure*}

However, we argue that tools fundamentally differ from ordinary text documents, making tool retrieval not merely a direct adaptation of traditional ad-hoc retrieval. 
Specifically, we identify three key limitations.
First, tool documentation is often incomplete and inconsistent in structures. 
Existing repositories lack standardized documentation schemas, leading to large variations in format, granularity, and terminology. 
Some documentation even omits critical functional details. 
For example, tools in Gorilla~\cite{gorilla} only provide parameter names without detailed descriptions.

Second, there are significant mismatches in semantics and granularity between user queries and tool documentation.
User queries are often ambiguous and expressed at a high level for a specific task, often involving coordinated use of multiple tools.
In contrast, tool documentation describes specific, atomic operations written in a technical manner.
Such discrepancies make ad-hoc retrieval models tailored for semantic and topical relevance matching not suitable for tool retrieval.

Third, more importantly, the utility of tools for a query is a multi-aspect concept rather than a simple measure of semantic similarity.
Beyond textual relevance, tool usefulness also depends on factors like parameter availability, execution feasibility, and downstream compatibility. 
These aspects differ substantially in importance and representation, making it difficult, if not impossible, to capture using a single relevance model. 
In our preliminary experiments, masking different fields in tool documentation leads to highly inconsistent impacts on retrieval performance across datasets (\cref{fig:mask_field_experiment}), suggesting applying a retrieval model on raw tool documentation is fundamentally suboptimal.

Based on these observations, we propose a \textbf{Multi-Field Tool Retrieval} framework (MFTR) for tool retrieval.
We first introduce a standardized schema to normalize heterogeneous tool documentation, mitigating the issue of documentation incompleteness.
We then design a tool-oriented query rewriting pipeline to align user intent with tool representations, addressing the semantic and granularity mismatch between queries and tools. 
Building upon this alignment process, MFTR performs fine-grained multi-field retrieval by modeling relevance separately for different fields.
An adaptive weighting mechanism further dynamically balances the contribution of each field to better model the actual utility of a tool.
We evaluate both our proposed method and SOTA baselines on five representative tool retrieval datasets separately, and also construct a new benchmark by mixing these datasets together to simulate a more realistic, large-scale, heterogeneous tool retrieval collection. 
Experimental results show that MFTR consistently improves retrieval performance across different retriever backbones in both single-dataset and mixed-dataset settings, highlighting the effectiveness and robustness of fine-grained multi-field relevance modeling.

In summary, the contributions of our paper are as follows:
\begin{itemize}[leftmargin=*, topsep=2pt, partopsep=0pt,
    before={\interlinepenalty=0 \clubpenalty=0 \widowpenalty=0}, itemsep=2pt]
    \item We identify limitations of treating tool documentation as flat text and reveal the variance in field contributions to retrieval performance.
    
    \item We propose MFTR, a comprehensive framework that introduces a standardized schema, query rewriting and alignment, and an adaptive multi-field weighting mechanism to precisely model tool utility and user intent.
    
    \item Experiments on multiple benchmarks demonstrate that MFTR significantly advances in tool retrieval and exhibits strong robustness across different retriever backbones.
\end{itemize}

\section{Related Work}
\label{sec:related_work}

\begin{figure*}[t]
    \centering
    \includegraphics[width=\linewidth]{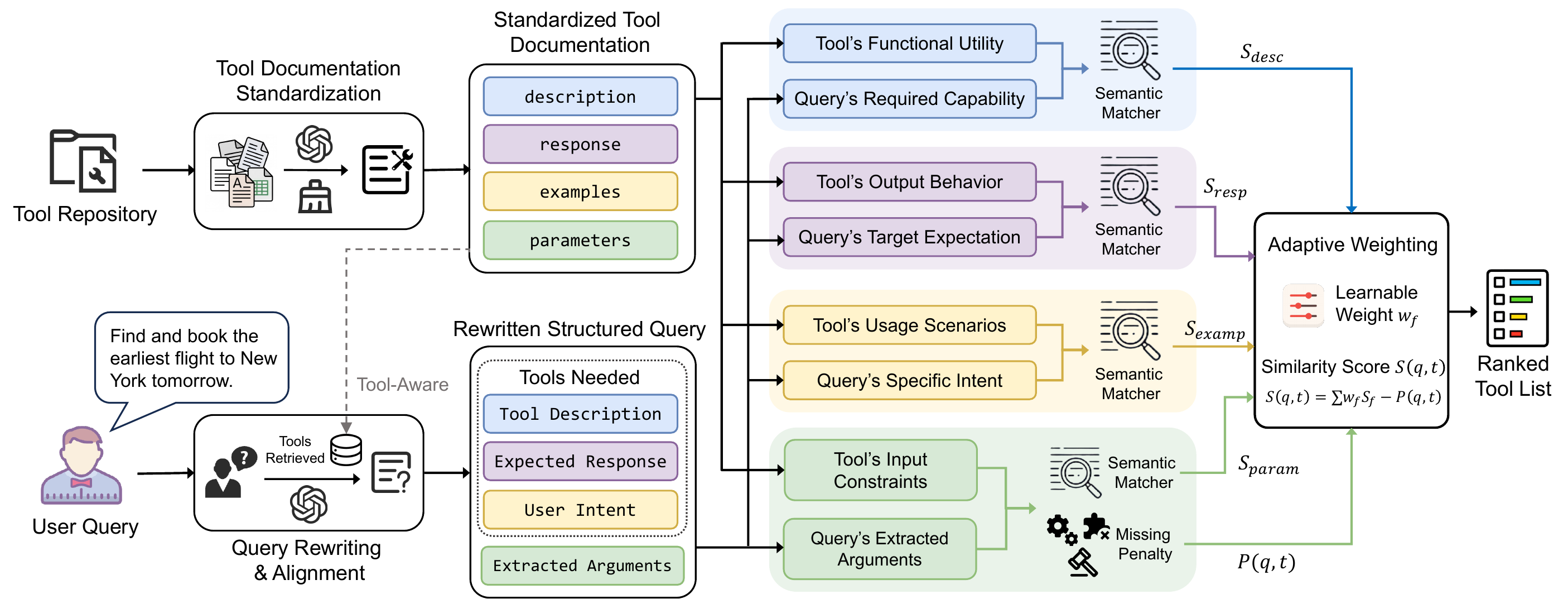}
    \caption{An illustration of our MFTR framework.}
    \label{fig:overall_framework}
\end{figure*}

To extend the capabilities of LLMs, prior work equips them with external tools via either tuning-free agentic interaction or tuning-based finetuning on specific tool use datasets~\cite{Mialon2023AugmentedLM, Qin2023ToolLW, 10.5555/3666122.3668004, 10.1609/aaai.v38i16.29759}.
Despite progress, scaling tool use to large and frequently updated repositories remains challenging due to context limits and continual-update costs, making tool retrieval a key prerequisite for practical tool use systems~\cite{colt, qu2025from}.
Existing tool retrieval approaches typically directly adopted traditional information retrieval methods, utilizing sparse retrievers (e.g., BM25~\cite{bm25}) or semantic-based dense retrieval models, sometimes enhanced by documentation refinement or feedback-driven embedding updates~\cite{yuan-etal-2025-easytool, lu2025toolsunderdocumentedsimpledocument, onlinerag}.
Others train tool-specific retrievers or adopt multi-step agentic workflows to improve recall and selection~\cite{toolbench, colt, pluto, shao-etal-2023-enhancing, braunschweiler-etal-2025-toolreagt}.
Despite these advances, existing methods remain within the ad-hoc retrieval paradigm, modeling the relevance between the query and the entire raw tool documentation.
This coarse-grained approach relies heavily on semantic similarity, overlooking the multi-aspect nature of tool utility.
In contrast, we propose independently modeling user queries and multi-field tool information for multi-dimensional alignment.

\section{Methodology}
\label{sec:methodology}

In this section, we present Multi-Field Tool Retrieval (MFTR), a framework that decomposes tool retrieval into fine-grained matching over multiple functional fields, as illustrated in \cref{fig:overall_framework}.
By standardizing the tool documentation (\cref{sec:standard_tool_documentation}), rewriting queries (\cref{sec:query_rewriting}) to match these fields, and adaptively weighting the resulting multi-field relevance scores (\cref{sec:adaptive_weight}), MFTR can match the actual utility of a tool with respect to a user query more precisely.

\subsection{Problem Formulation and Overview}
\label{sec:problem_formulation_and_overview}

Let $\mathcal{T} = \{(t_1, d_1), (t_2, d_2), \cdots, (t_N, d_N) \}$ denote a tool repository, where each tool $t_i$ is associated with documentation $d_i$.
Given a query $q$, tool retrieval aims to return a relevant subset $\mathcal{T}_q \subset \mathcal{T}$.
Conventional retrievers score a tool by treating $d_i$ as an unstructured text and compute a semantic similarity score by a relevance function $\phi(\cdot, \cdot)$:
\begin{equation}
    S(q, t_i) = \phi(q, d_i),
\end{equation}
However, this paradigm neglects query-tool mismatch and the varying importance of different functional fields.
MFTR addresses this by representing tools and queries with aligned functional fields and aggregating per-field relevance.

\subsection{Tool Documentation Standardization}
\label{sec:standard_tool_documentation}

As discussed in \cref{sec:introduction}, tool documentation from different sources is often inconsistent in structure, terminology, and granularity, as shown in the examples in~\cref{appendix:tool_document_case}.
Furthermore, identical concepts are often named with inconsistent terminology across datasets (e.g., \texttt{parameters} vs. \texttt{api\_arguments}).
Such heterogeneity makes it difficult for retrievers to align tools from different repositories into a shared semantic space.
To address these issues, we introduce a tool documentation standardization module that transforms heterogeneous raw documentation into a unified structured schema using an LLM.
The core challenge in designing a standardized schema lies in determining the optimal set of fields.
Our selection is guided by two objectives:

\begin{itemize}[leftmargin=*]

    \item \textbf{Functional Coverage.} 
    The selected fields should collectively capture the critical functional aspects required to understand and use a tool, including its purpose, inputs, outputs, and typical usage scenarios.
    Insufficient functional coverage would otherwise lead to false positives, where a tool appears semantically relevant but is practically unusable.

    \item \textbf{Generalization and Robustness.} 
    The schema should remain minimal to ensure generalizability and reliability. 
    Defining too many unmentioned fields may lead the LLM to hallucinate details. 
    Furthermore, some fields provide little retrieval-relevant information and may even introduce noise.
\end{itemize}

Balancing coverage and generalizability, we define a standardized schema with four fields: \texttt{description}, \texttt{parameters}, \texttt{response}, and \texttt{examples}.
These fields correspond to a tool’s functionality, inputs, outputs, and usage scenarios, respectively, which are widely present in the raw tool documentation but often expressed in unstructured or inconsistent forms.
To construct the standardized documentation, We prompt an LLM to extract and reformat the information for each field from the raw documentation.
During extraction, unless the raw documentation explicitly marks a parameter as optional, we mark it as required.
This standardization module provides a clean and informative foundation for subsequent multi-field representation and matching.

\subsection{Query Rewriting and Alignment}
\label{sec:query_rewriting}

As highlighted in \cref{sec:introduction}, a fundamental challenge in tool retrieval is semantic and granularity mismatch: user queries are often high-level, ambiguous, and multi-intent, whereas tools are technically precise and atomic.
To bridge this mismatch, we rewrite user queries into a structured representation that aligns with our standardized tool documentation schema.
Formally, given a user query $q$, since $q$ may involve multiple tool needs, we first decompose it into a sequence of tool needs $\{a_1, a_2, \cdots, a_k\}$, each corresponding to a functional intent implied by the query. For each tool need $a_i$, we define three aligned sub-fields:

\begin{itemize}[leftmargin=*]
    \item \textbf{User Intent} (aligned with \texttt{examples}): A concrete user utterance objective of the tool need.
    
    \item \textbf{Tool Description} (aligned with \texttt{description}): A formal technical summary of the required functionality.

    \item \textbf{Expected Response} (aligned with \texttt{response}): An abstract specification of the anticipated output content and format required by the user's request.
\end{itemize}

\noindent In addition, we extract \textbf{Extracted Arguments} (aligned with the \texttt{parameters} field) to identify the semantic roles and data types of all parameters mentioned or implied in $q$.

\paragraph{Rewriting via Tool-Aware Pseudo-Relevance Feedback.}
We argue that an effective tool-using system must maintain explicit awareness of its available tools to make informed decisions.
Similarly, the query rewriting module should also be aware of the available tool repository to adopt the consistent terminology and technical expressions used in the tool documentation.
To this end, we leverage the Pseudo-Relevance Feedback paradigm to inject knowledge of the tool repository into the rewriting process. 
Specifically, we treat the standardized descriptions generated in \cref{sec:standard_tool_documentation} as a corpus, retrieve the top-$K$ (e.g., $K=20$) most relevant descriptions to $q$ using BM25~\cite{bm25}, and feed them as pseudo-positive examples.
In addition, we include the full documentation of the most relevant tools to help the LLM understand the exact schema of tool documentation.
Conditioned on this pseudo-relevant information, the LLM rewrites the query into the predefined structured format, effectively aligning the generation process with the semantic space of the target tools.
By rewriting queries into a field-aligned structured form, MFTR is able to independently model each specific field and optimize retrieval, facilitating more precise and robust tool selection.

\subsection{Adaptive Weighting}
\label{sec:adaptive_weight}

Based on the standardized tool documentation and the rewritten structured query, we propose a multi-field relevance computation module to achieve precise retrieval considering tool utility.
Given a tool $t$ and a user query $q$, let $d'$ denote the standardized tool documentation and $q'$ denote the rewritten structured query.
For convenience, we use the tool documentation schema to name the format of $d'$ and $q'$, which means they all consist of four aligned fields: \texttt{description}, \texttt{parameters}, \texttt{response}, and \texttt{examples}.
MFTR computes field-wise relevance and aggregates them into a unified ranking score.

\subsubsection{Semantic Field Matching}

For semantic intent fields \texttt{description}, \texttt{response} and \texttt{examples}, we score each field using a retriever.
Since a single user query may be decomposed into $k$ tool needs during the rewriting phase, a tool is considered a match if it satisfies any of these needs.
Formally, the similarity score $S_f(q, t)$ for a field $f$ is defined as the maximum similarity across all tool needs:
\begin{equation}
    \begin{aligned}
        & \ \ \ S_f(q, t) = \max_{i=1}^k \phi ({q'}_f^i, {d'}_f), \\
        \forall f \in  &\{\small\texttt{description, response, examples}\}.
    \end{aligned}
\end{equation}
where ${q'}_f^i$ is the field $f$ of $i$-th tool need, ${d'}_f$ is the field $f$ in the tool documentation, and 
$\phi(\cdot, \cdot)$ denotes a relevance function instantiated by the underlying retriever.

\subsubsection{Parameter Alignment with Missing Penalty}
\label{sec:soft_penalty}

Unlike other fields that measure intent-level similarity, the \texttt{parameters} field requires functional constraints: a tool is executable if and only if all parameters are provided or can be inferred.
To model parameter completeness, we introduce an adaptive penalty that detects poorly matched parameters and applies penalties based on their importance (required vs. optional).

Let $\mathcal{P}_t$ be tool $t$'s parameters and $\mathcal{A}_q$ denote the arguments extracted from the rewritten query.
For each parameter $p_j$, the best-matching argument score is:
\begin{equation}
    s_{p_j} = \max_{a \in \mathcal{A}_q} \phi(p_j, a), \forall p_j \in \mathcal{P}_t
\end{equation}
The base parameter matching score is the average:
\begin{equation}
    S_{\texttt{param}}(q, t) = \frac{1}{|\mathcal{P}_t|} \sum_{p_j \in \mathcal{P}_t} s_{p_j}
\end{equation}
\indent We use an adaptive detection mechanism to identify "missing" parameters based on matching scores.
Since similarity scores from different retrievers (e.g., BM25 and dense retrievers) differ substantially in scale and distribution, a fixed cutoff is suboptimal.
Instead, we introduce a learnable parameter $\tau$ that acts as a normalization threshold, adapting to the specific score distribution of the underlying retriever.
The penalty of the parameter $p_j$ is defined using a sigmoid-based smoothed indicator function:
\begin{equation}
{\fontsize{10.3}{11.5}\selectfont
\begin{aligned}
L(p_j) &= \frac{1}{1+\exp(\alpha (\tau - s_{p_j}))} \\
&\quad \cdot \Bigl(
w_{\text{req}} \mathbb{I}_{\text{req}}(p_j)
+ w_{\text{opt}} \bigl(1 - \mathbb{I}_{\text{req}}(p_j)\bigr)
\Bigr)
\end{aligned}}
\end{equation}
where $\mathbb{I}_{\text{req}}$ marks required parameters,
$w_{\text{req}}$ and $w_{\text{opt}}$ are learnable penalty weights for required and optional parameters, respectively.
$\alpha$ is a hyperparameter that controls the shape of the sigmoid curve.
Intuitively, when $s_{p_j} < \tau$, the sigmoid term approaches $1$, activating the penalty; otherwise, it vanishes smoothly.

The total parameter penalty $P(q, t)$ is the sum of penalties over all parameters:
\begin{equation}
    P(q, t) = \sum_{p_j \in \mathcal{P}_t} L(p_j)
\end{equation}
This mechanism ensures that tools missing key required parameters receive lower ranking scores, enforcing functional correctness.


\subsubsection{Adaptive Aggregation and Optimization}
\label{sec:training_weights}

We aggregate multi-field similarities with the parameter penalty into a final relevance score as a linear combination:
\begin{equation}
    S(q, t) = \sum_{f} w_f \cdot S_f(q, t) + b - P(q, t)
\end{equation}
where $w_f$ are learnable field weights, and $b$ is the bias term.
We optimize using a pairwise ranking loss~\cite{ranknet}. 
For a positive tool $t^+$ and a negative tool $t^-$ given query $q$, the objective is to maximize the margin between the scores of positive and negative samples by minimizing the following loss:
\begin{equation}
    \mathcal{L} = \log \Bigl( 1 + \exp \bigl( - (S(q, t^+) - S(q, t^-)) \bigr) \Bigr)
\end{equation}
By end-to-end training, MFTR adaptively learns the importance of different functional fields and the rigorousness of parameter constraints.

\subsection{Discussion}
\label{sec:discussion_retrieval_vs_execution}

It is worth noting that the standardized tool documentation and rewritten structured queries in MFTR are designed specifically for retrieval rather than direct tool execution.
In practical deployment, we recommend using the original tool documentation, as it preserves complete implementation details and avoids hallucination introduced during standardization.
As long as we use the original tool documentation for actual tool use, LLM-based reformulation in MFTR would not affect the usability of the tool.

\section{Experimental Setup}
\label{sec:experimental_setup}

\subsection{Datasets and Evaluation Metrics}
\label{sec:datasets_and_evaluation_metrics}

We evaluate on five tool retrieval benchmarks adapted from tool learning datasets: ToolBench~\cite{toolbench}, APIGen~\cite{apigen}, APIBank~\cite{apibank}, Gorilla (HuggingFace subset)~\cite{gorilla}, and Toolink~\cite{toolink}.
We also construct a \textbf{Mixed} benchmark by merging all five datasets to assess robustness under a large, heterogeneous tool pool.
Details and statistics of these datasets are provided in~\cref{appendix:datasets}.
We evaluate retrieval performance using two standard IR metrics: (i) NDCG@K (N@K), which considers both the relevance and ranking positions; and (ii) Recall@K (R@K) evaluates the proportion of relevant tools that appear within the top-K retrieved results.
We further provide experiments extending MFTR to agent skill retrieval in~\cref{appendix:skill_retrieval}.

\subsection{Retrieval Models}

We experiment with various retrievers.
For sparse retrieval, we consider BM25~\cite{bm25}.
For dense retrieval, we evaluate six representative models spanning different sizes: all-MiniLM-L6-v2 (MiniLM-L6), Contriever~\cite{contriever} trained on MS-MARCO~\cite{msmarco}, multilingual-e5-base and multilingual-e5-large (E5-Base/Large)~\cite{multilingual_e5}, gte-large-en-v1.5 (GTE)~\cite{gte_retriever}, and bge-large-en-v1.5 (BGE)~\cite{bge_retriever}.
We further compare two retrieval models specifically trained for ad-hoc tool retrieval: API Retriever~\footnote{\url{https://huggingface.co/ToolBench/ToolBench_IR_bert_based_uncased}}~\cite{toolbench} (based on BERT~\cite{Devlin2019BERT}) and COLT~\footnote{\url{https://huggingface.co/Tool-COLT/contriever-base-msmarco-v1-ToolBenchG3}}~\cite{colt} (based on Contriever).

\subsection{Baselines and Implementation Details}

We compare MFTR against four representative baselines:
(1) \textbf{Full-Doc} serves as the basic semantic retrieval on raw documentation.
(2) \textbf{EasyTool}~\cite{yuan-etal-2025-easytool} represents the documentation enhancement approach, which uses LLMs to enrich tool descriptions and guidelines.
(3) \textbf{PLUTo}~\cite{pluto} includes the query-centric paradigm, employing query decomposition and an Edit-and-Ground pipeline for tool documentation.
(4) \textbf{OnlineRAG}~\cite{onlinerag} is one of the latest approaches, which introduces a feedback-driven mechanism to continuously refine tool representations. Note that OnlineRAG requires updating tool embeddings, so we do not apply it when using BM25 as the retriever.

We adopt gpt-4o-mini for documentation standardization and query rewriting (we also tested Qwen3.5-35B-A3B in~\cref{appendix:ablation_llm}), and conduct 5-fold cross-validation for robust evaluation. 
The multi-field weighting model is trained for 5 epochs with BM25-based negative sampling. Detailed settings are provided in \cref{appendix:baselines,appendix:implementation_details}.

\section{Results and Analysis}
\label{sec:results}

\begin{table*}[t]
  \small
  \centering
  \caption{
    Overall experimental results on five datasets and the Mixed benchmark.
    We report NDCG@10 (N@10) and Recall@10 (R@10).
    COLT~\cite{colt} and API Retriever~\cite{toolbench} are specialized tool retrieval models trained on ToolBench~\cite{toolbench}.
    \textbf{Bold} and \underline{underlined} values indicate the best and second-best results, respectively.
    ``${\dagger}$'' denotes that the improvement of our method over the strongest baseline is statistically significant under the Fisher’s Randomization Test with $p < 0.05$.
    Note that OnlineRAG requires updating tool embeddings, so we do not apply it when using BM25 as the retriever.
  }
  \resizebox{\textwidth}{!}
  {
    \begin{tabular}{c|c||cc|cc|cc|cc|cc||cc}
    \toprule    \multirow{2}{*}[-1.5pt]{\textbf{Model}} & \multirow{2}{*}[-1.5pt]{\textbf{Method}} & \multicolumn{2}{c|}{\textbf{ToolBench}} & \multicolumn{2}{c|}{\textbf{APIGen}} & \multicolumn{2}{c|}{\textbf{APIBank}} & \multicolumn{2}{c|}{\textbf{Gorilla}} & \multicolumn{2}{c||}{\textbf{Toolink}} & \multicolumn{2}{c}{\textbf{Mixed}} \\
\cmidrule{3-14}          &       & \textbf{N@10} & \textbf{R@10} & \textbf{N@10} & \textbf{R@10} & \textbf{N@10} & \textbf{R@10} & \textbf{N@10} & \textbf{R@10} & \textbf{N@10} & \textbf{R@10} & \textbf{N@10} & \textbf{R@10} \\
    \toprule
    \multirow{5}[1]{*}{BM25} & Full-Doc & 0.3700 & 0.4306 & 0.7571 & 0.8579 & 0.5167 & 0.6139 & 0.2379 & 0.3520 & 0.3828 & 0.4423 & 0.3947 & 0.4858 \\
          & EasyTool & \underline{0.5245} & \underline{0.6015} & \underline{0.7928} & \underline{0.8882} & \underline{0.5637} & \underline{0.6551} & \underline{0.2697} & \underline{0.4220} & \underline{0.4899} & \underline{0.5369} & \underline{0.4668} & \underline{0.5869} \\
          & PLUTo & 0.2844 & 0.3754 & 0.5328 & 0.7392 & 0.4222 & 0.5033 & 0.1852 & 0.3180 & 0.3661 & 0.4001 & 0.4667 & 0.5715 \\
          & OnlineRAG & -     & -     & -     & -     & -     & -     & -     & -     & -     & -     & -     & - \\
          & MFTR (ours) & \textbf{0.5573$^{\dagger}$} & \textbf{0.6270$^{\dagger}$} & \textbf{0.8430$^{\dagger}$} & \textbf{0.9228$^{\dagger}$} & \textbf{0.6704$^{\dagger}$} & \textbf{0.6906} & \textbf{0.3716$^{\dagger}$} & \textbf{0.5620$^{\dagger}$} & \textbf{0.5497$^{\dagger}$} & \textbf{0.6200$^{\dagger}$} & \textbf{0.5807$^{\dagger}$} & \textbf{0.6626$^{\dagger}$} \\
    \midrule
    \multirow{5}[0]{*}{MiniLM-L6} & Full-Doc & 0.2790 & 0.3432 & 0.7126 & 0.8178 & \underline{0.6112} & \underline{0.7442} & 0.1808 & 0.3100 & 0.3818 & 0.4308 & 0.3273 & 0.4178 \\
          & EasyTool & \underline{0.4306} & \underline{0.5122} & 0.7193 & \underline{0.8359} & 0.6109 & 0.7393 & 0.2336 & 0.3780 & 0.4279 & 0.4599 & 0.3109 & 0.4145 \\
          & PLUTo & 0.3486 & 0.4407 & 0.5680 & 0.7181 & 0.5218 & 0.5767 & 0.1843 & 0.2900 & 0.3926 & 0.4450 & \underline{0.3895} & \underline{0.4931} \\
          & OnlineRAG & 0.4139 & 0.4821 & \underline{0.7240} & 0.8281 & 0.2016 & 0.2962 & \underline{0.2582} & \underline{0.3960} & \underline{0.4590} & \underline{0.5258} & 0.3445 & 0.4511 \\
          & MFTR (ours) & \textbf{0.5412$^{\dagger}$} & \textbf{0.6256$^{\dagger}$} & \textbf{0.8516$^{\dagger}$} & \textbf{0.9248$^{\dagger}$} & \textbf{0.7237$^{\dagger}$} & \textbf{0.7863$^{\dagger}$} & \textbf{0.3775$^{\dagger}$} & \textbf{0.5860$^{\dagger}$} & \textbf{0.5334$^{\dagger}$} & \textbf{0.6070$^{\dagger}$} & \textbf{0.5617$^{\dagger}$} & \textbf{0.6449$^{\dagger}$} \\
    \midrule
    \multirow{5}[0]{*}{Contriever} & Full-Doc & 0.3601 & 0.4221 & 0.7096 & 0.8178 & 0.5894 & \underline{0.6898} & 0.2374 & 0.3760 & 0.3809 & 0.4026 & 0.3762 & 0.4608 \\
          & EasyTool & 0.4593 & \underline{0.5462} & 0.7361 & 0.8355 & \underline{0.5960} & 0.6815 & \underline{0.2720} & \underline{0.4200} & \underline{0.5052} & \underline{0.5268} & 0.4199 & 0.5287 \\
          & PLUTo & 0.3445 & 0.4320 & 0.5740 & 0.7242 & 0.4859 & 0.5371 & 0.1878 & 0.2960 & 0.3859 & 0.4096 & \underline{0.4396} & \underline{0.5439} \\
          & OnlineRAG & \underline{0.4677} & 0.5373 & \underline{0.7532} & \underline{0.8539} & 0.2177 & 0.3366 & 0.2516 & 0.3840 & 0.3455 & 0.4992 & 0.3831 & 0.4761 \\
          & MFTR (ours) & \textbf{0.5275$^{\dagger}$} & \textbf{0.6103$^{\dagger}$} & \textbf{0.8365$^{\dagger}$} & \textbf{0.9127$^{\dagger}$} & \textbf{0.6816$^{\dagger}$} & \textbf{0.7203} & \textbf{0.3823$^{\dagger}$} & \textbf{0.6000$^{\dagger}$} & \textbf{0.5493$^{\dagger}$} & \textbf{0.6137$^{\dagger}$} & \textbf{0.5624$^{\dagger}$} & \textbf{0.6499$^{\dagger}$} \\
    \midrule
    \multirow{5}[0]{*}{E5-Base} & Full-Doc & 0.4722 & 0.5440 & 0.7359 & 0.8358 & \underline{0.6190} & \underline{0.7442} & 0.2667 & 0.4140 & 0.3779 & 0.4805 & 0.4109 & 0.4928 \\
          & EasyTool & \underline{0.5309} & \underline{0.6085} & \underline{0.7538} & \underline{0.8585} & 0.6148 & 0.6873 & 0.2867 & 0.4300 & \underline{0.4219} & 0.5047 & 0.4018 & 0.5036 \\
          & PLUTo & 0.3315 & 0.4293 & 0.5534 & 0.7285 & 0.4957 & 0.5817 & 0.1665 & 0.2720 & 0.3263 & 0.3584 & \underline{0.4555} & \underline{0.5628} \\
          & OnlineRAG & 0.5265 & 0.5928 & 0.5140 & 0.8556 & 0.2216 & 0.3094 & \underline{0.2965} & \underline{0.4440} & 0.3676 & \underline{0.5402} & 0.4186 & 0.5085 \\
          & MFTR (ours) & \textbf{0.5520$^{\dagger}$} & \textbf{0.6295$^{\dagger}$} & \textbf{0.8516$^{\dagger}$} & \textbf{0.9263$^{\dagger}$} & \textbf{0.7320$^{\dagger}$} & \textbf{0.7632} & \textbf{0.3575$^{\dagger}$} & \textbf{0.5520$^{\dagger}$} & \textbf{0.5314$^{\dagger}$} & \textbf{0.5936$^{\dagger}$} & \textbf{0.5806$^{\dagger}$} & \textbf{0.6570$^{\dagger}$} \\
    \midrule
    \multirow{5}[0]{*}{E5-Large} & Full-Doc & 0.4577 & 0.5205 & 0.7442 & 0.8430 & \underline{0.6412} & \underline{0.7393} & 0.2729 & 0.3980 & \underline{0.3992} & 0.4524 & 0.4292 & 0.5155 \\
          & EasyTool & \underline{0.5526} & \underline{0.6281} & \underline{0.7774} & 0.8708 & 0.6373 & 0.7195 & \underline{0.3131} & \underline{0.4940} & 0.3824 & 0.4348 & 0.4168 & 0.5253 \\
          & PLUTo & 0.3420 & \multicolumn{1}{c}{0.4094} & 0.5597 & 0.7325 & 0.5304 & 0.5982 & 0.1655 & 0.2700 & 0.3012 & 0.3396 & \underline{0.4746} & \underline{0.5835} \\
          & OnlineRAG & 0.5346 & 0.6028 & 0.5282 & \underline{0.8815} & 0.2732 & 0.4414 & 0.3073 & 0.4540 & 0.3542 & \underline{0.5104} & 0.4291 & 0.5309 \\
          & MFTR (ours) & \textbf{0.5606} & \textbf{0.6359} & \textbf{0.8556$^{\dagger}$} & \textbf{0.9302$^{\dagger}$} & \textbf{0.7219$^{\dagger}$} & \textbf{0.7426} & \textbf{0.3551$^{\dagger}$} & \textbf{0.5540$^{\dagger}$} & \textbf{0.5796$^{\dagger}$} & \textbf{0.6509$^{\dagger}$} & \textbf{0.5856$^{\dagger}$} & \textbf{0.6700$^{\dagger}$} \\
    \midrule
    \multirow{5}[0]{*}{GTE-Large} & Full-Doc & 0.4444 & 0.5221 & 0.6931 & 0.8061 & 0.6331 & 0.6972 & 0.3236 & 0.5040 & \underline{0.3914} & 0.4272 & 0.3869 & 0.4916 \\
          & EasyTool & \underline{0.5138} & \underline{0.5984} & \underline{0.7577} & \underline{0.8619} & \underline{0.6629} & \textbf{0.7343} & \underline{0.3598} & \underline{0.5520} & 0.3387 & 0.3749 & 0.4360 & 0.5435 \\
          & PLUTo & 0.3474 & 0.4567 & 0.5600 & 0.7094 & 0.4971 & 0.5644 & 0.1995 & 0.3160 & 0.3761 & 0.4184 & \underline{0.4595} & \underline{0.5684} \\
          & OnlineRAG & 0.4294 & 0.5135 & 0.6984 & 0.8145 & 0.2406 & 0.3350 & 0.3294 & 0.4840 & 0.3672 & \underline{0.5426} & 0.3624 & 0.4668 \\
          & MFTR (ours) & \textbf{0.5348$^{\dagger}$} & \textbf{0.6111} & \textbf{0.8476$^{\dagger}$} & \textbf{0.9272$^{\dagger}$} & \textbf{0.6888} & \underline{0.7195} & \textbf{0.3849} & \textbf{0.5980$^{\dagger}$} & \textbf{0.5602$^{\dagger}$} & \textbf{0.6130$^{\dagger}$} & \textbf{0.5634$^{\dagger}$} & \textbf{0.6553$^{\dagger}$} \\
    \midrule
    \multirow{5}[1]{*}{BGE-Large} & Full-Doc & 0.3982 & 0.4794 & 0.7748 & 0.8662 & \underline{0.6856} & \textbf{0.7525} & 0.2074 & 0.3220 & 0.4213 & 0.4805 & 0.3853 & 0.4778 \\
          & EasyTool & \underline{0.4900} & \underline{0.5657} & 0.7355 & 0.8408 & 0.6249 & 0.7393 & \underline{0.3035} & \underline{0.4560} & \underline{0.4643} & 0.5188 & 0.4353 & 0.5382 \\
          & PLUTo & 0.3587 & 0.4814 & 0.5634 & 0.7217 & 0.5557 & 0.6163 & 0.1521 & 0.2340 & 0.3490 & 0.3901 & \underline{0.4530} & \underline{0.5546} \\
          & OnlineRAG & 0.4780 & 0.5466 & \underline{0.7794} & \underline{0.8835} & 0.2350 & 0.3614 & 0.2878 & 0.4360 & 0.4232 & \underline{0.6325} & 0.4038 & 0.5008 \\
          & MFTR (ours) & \textbf{0.5544$^{\dagger}$} & \textbf{0.6347$^{\dagger}$} & \textbf{0.8575$^{\dagger}$} & \textbf{0.9308$^{\dagger}$} & \textbf{0.7170} & \underline{0.7459} & \textbf{0.3796$^{\dagger}$} & \textbf{0.5880$^{\dagger}$} & \textbf{0.5725$^{\dagger}$} & \textbf{0.6489} & \textbf{0.5900$^{\dagger}$} & \textbf{0.6811$^{\dagger}$} \\

    \midrule

    \multirow{5}[0]{*}{COLT} & Full-Doc & 0.4110 & 0.4905 & 0.7846 & 0.8890 & 0.5441 & 0.6625 & 0.2160 & 0.3500 & 0.4023 & 0.4276 & 0.4063 & 0.4966 \\
          & EasyTool & \underline{0.5378} & \textbf{0.6351} & \underline{0.7953} & \underline{0.8900} & \underline{0.6271} & \underline{0.7393} & \underline{0.2393} & \underline{0.3740} & \underline{0.5378} & \underline{0.5922} & 0.4374 & 0.5454 \\
          & PLUTo & 0.3536 & 0.4876 & 0.5254 & 0.7464 & 0.5055 & 0.6502 & 0.1270 & 0.2180 & 0.3631 & 0.4137 & 0.4568 & \underline{0.5602} \\
          & OnlineRAG & \textbf{0.5437} & \underline{0.6253} & 0.7739 & 0.8765 & 0.2296 & 0.3342 & 0.2122 & 0.3240 & 0.1994 & 0.2341 & \underline{0.4579} & 0.5549 \\
          & MFTR (ours) & 0.5011 & 0.5836 & \textbf{0.8460$^{\dagger}$} & \textbf{0.9216$^{\dagger}$} & \textbf{0.6456} & \textbf{0.7401} & \textbf{0.3764$^{\dagger}$} & \textbf{0.5660$^{\dagger}$} & \textbf{0.5479} & \textbf{0.6214$^{\dagger}$} & \textbf{0.5482$^{\dagger}$} & \textbf{0.6385$^{\dagger}$} \\

    \midrule
          
    \multirow{5}[0]{*}{API Retriever} & Full-Doc & 0.4944 & 0.5703 & 0.7270 & 0.8215 & 0.5171 & 0.6551 & 0.1607 & 0.2740 & 0.2727 & 0.3618 & 0.3861 & 0.4756 \\
          & EasyTool & \underline{0.5515} & \underline{0.6203} & \underline{0.7420} & \underline{0.8315} & \underline{0.5603} & \underline{0.6972} & \underline{0.1988} & \underline{0.3140} & \underline{0.3057} & \underline{0.4311} & 0.3946 & 0.5055 \\
          & PLUTo & 0.3934 & 0.5461 & 0.5392 & 0.7482 & 0.4863 & 0.5817 & 0.1024 & 0.1700 & 0.2954 & 0.3161 & \underline{0.4173} & \underline{0.5102} \\
          & OnlineRAG & 0.5424 & 0.6115 & 0.7367 & 0.8274 & 0.2346 & 0.3284 & 0.1789 & 0.2900 & 0.3052 & 0.3850 & 0.3776 & 0.4743 \\
          & MFTR (ours) & \textbf{0.5854$^{\dagger}$} & \textbf{0.6743$^{\dagger}$} & \textbf{0.8364$^{\dagger}$} & \textbf{0.9163$^{\dagger}$} & \textbf{0.6923$^{\dagger}$} & \textbf{0.7838$^{\dagger}$} & \textbf{0.3330$^{\dagger}$} & \textbf{0.5240$^{\dagger}$} & \textbf{0.4964$^{\dagger}$} & \textbf{0.5493$^{\dagger}$} & \textbf{0.5582$^{\dagger}$} & \textbf{0.6493$^{\dagger}$} \\

    \bottomrule
    \end{tabular}%
  }
  \label{tab:overall_results}%
\end{table*}%

In this section, we evaluate MFTR. 
We first compare with baselines across datasets and retrievers. 
Then we ablate each component. 
Finally, we analyze the impact of different documentation fields and discuss efficiency.
Additional results and analyses are provided in~\cref{appendix:experiments}.

\subsection{Main Results}
\label{sec:main_results}

\cref{tab:overall_results} shows MFTR consistently outperforming baselines across five datasets and the Mixed benchmark, with statistically significant gains in many cases.
It clearly surpasses the Full-Doc paradigm, supporting our claim that tool retrieval is not a simple extension of ad-hoc retrieval and that flattening documentation is insufficient.
Compared to strong baselines like EasyTool and OnlineRAG, MFTR maintains a clear lead, highlighting the critical role of multi-field modeling in capturing fine-grained semantic features.
As a model-agnostic framework, MFTR consistently improves retrievers of different architectures, including specialized ones (COLT and API Retriever).
Notably, MiniLM-L6 with MFTR matches or surpasses models with 10$\times$ more parameters, suggesting gains mainly come from multi-field relevance modeling rather than model size.

\begin{figure}[!t]
    \centering    
    \includegraphics[width=\linewidth]{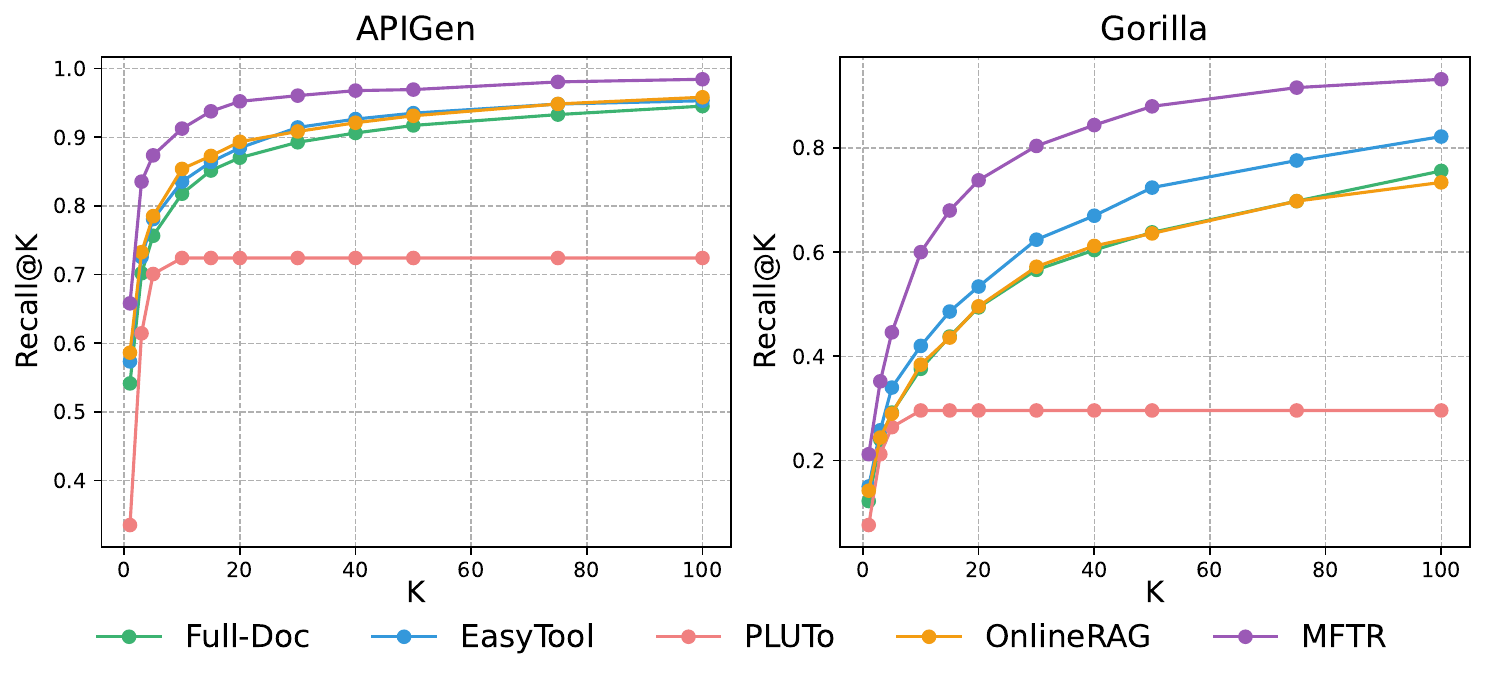}
    \caption{Recall@K across different $K$ using Contriever.}
    \label{fig:recall_curve}
\end{figure}

MFTR shows remarkable adaptability to datasets of varying quality.
On Gorilla, where documentation suffers from redundancy and missing parameters, MFTR reconstructs useful structure; with MiniLM-L6, it improves Recall@10 from 0.3960 to 0.5860 (+47.98\%)
On APIGen with high-quality documentation, MFTR achieves an 11.42\% relative NDCG@10 gain over the runner-up, indicating that even in high-quality retrieval environments, fine-grained modeling captures signals missed by coarse matching.
OnlineRAG underperforms on smaller datasets (e.g., APIBank) due to limited feedback, whereas MFTR remains strong without large-scale training.
Across retrieval depths, \cref{fig:recall_curve} shows MFTR outperforms all baselines at lower $K$ values and consistently maintains this lead as $K$ increases.
Notably, PLUTo employs an LLM to select a limited number of tools from the retrieved candidates, so its recall remains constant as $K$ grows large.
On the Mixed benchmark, MFTR shows strong generalization.
It improves NDCG@10 and Recall@10 by 28.61\% and 18.98\% over the runner-up, respectively, indicating that standardizing tools and queries improves cross-source robustness.

\subsection{Downstream Tool Use Performance}
\label{sec:tool_utility}

\begin{table}[t]
  \centering
  \small
  \caption{
    Retrieval and downstream tool use performance on StableToolBench~\cite{guo-etal-2024-stabletoolbench}.
    SoPR denotes Solvable Pass Rate, reported as mean $\pm$ standard deviation
    over three evaluator runs. Best results within each retriever backbone are highlighted in
    \textbf{bold}.
  }
  \resizebox{\linewidth}{!}
  {
    \begin{tabular}{llccc}
    \toprule
    \textbf{Retriever} & \textbf{Method} & \textbf{N@10} & \textbf{R@10} & \textbf{SoPR (\%)} \\
    \midrule
    
    \multirow{1.75}[1]{*}{BM25} & Full-Doc & 0.3700 & 0.4306 & 50.5$\pm$0.5 \\
         & MFTR & \textbf{0.5573} & \textbf{0.6270} & \textbf{56.5$\pm$0.8} \\
    \midrule
    \multirow{1.75}[1]{*}{MiniLM-L6} & Full-Doc & 0.2790 & 0.3432 & 47.3$\pm$0.4 \\
              & MFTR & \textbf{0.5412} & \textbf{0.6256} & \textbf{59.2$\pm$0.5} \\
    \midrule
    \multirow{1.75}[1]{*}{Contriever} & Full-Doc & 0.3601 & 0.4221 & 50.9$\pm$0.4 \\
               & MFTR & \textbf{0.5275} & \textbf{0.6103} & \textbf{56.8$\pm$0.3} \\
    \midrule
    \multirow{1.75}[1]{*}{BGE-Large} & Full-Doc & 0.3982 & 0.4794 & 54.2$\pm$0.5 \\
              & MFTR & \textbf{0.5544} & \textbf{0.6347} & \textbf{59.7$\pm$0.8} \\
    \midrule
    \multirow{1.75}[1]{*}{API Retriever} & Full-Doc & 0.4944 & 0.5703 & 49.3$\pm$0.7 \\
                 & MFTR & \textbf{0.5854} & \textbf{0.6743} & \textbf{58.6$\pm$0.5} \\
    \bottomrule
    \end{tabular}%
  }
  \label{tab:downstream_tool_use}%
\end{table}

Although our primary focus is tool retrieval, we further evaluate whether retrieval improvements translate into downstream tool use performance on StableToolBench~\cite{guo-etal-2024-stabletoolbench}.
StableToolBench extends ToolBench with a virtual API server and a stable evaluation system to reduce the impact of API instability during tool use evaluation.
For each query, we retrieve the top-10 tools using either Full-Doc or MFTR and provide them to the same Qwen3-32B tool use agent.
We use gpt-4o-mini as the evaluator and report the Solvable Pass Rate (SoPR), averaged over three evaluator runs.
All other agent and execution settings are kept identical across the two retrieval methods.
The results are shown in \cref{tab:downstream_tool_use}.
MFTR consistently improves SoPR over Full-Doc across all five retriever backbones, with absolute gains ranging from 5.5 to 11.9 percentage points.
These gains are consistent with the improvements in NDCG@10 and Recall@10, indicating that MFTR's retrieval improvements can translate into better downstream task completion.

\subsection{Ablation Study}
\label{sec:ablation_study}

To investigate component contributions, we conduct an ablation study, examining Tool Documentation Standardization (Stand.), Query Rewriting (Rewrite), Adaptive Weighting, and Missing Penalty.
\textit{w/o Adaptive Weighting} denotes a multi-field baseline where the final relevance score is calculated as the mean of similarity scores of all fields.
Results are in \cref{tab:exp_ablation}.
We observe three key findings:
(1) Only \textit{Stand.} or \textit{Rewrite} used as textual augmentation yield limited and sometimes negative gains (e.g., APIGen), suggesting that naive LLM augmentation may introduce both information and noise.
(2) Standardization and query--tool structural alignment are essential: \textit{Weighting + Stand.} uses original query for matching fields showing good improvements, whereas \textit{Weighting + Rewrite} fails due to mismatched structured queries and raw documentation; \textit{Weighting Only} is also unstable, indicating that adaptive weighting cannot compensate for heterogeneous raw fields.
When both Stand. and Rewrite are enabled, independently modeled fields already bring substantial gains even under simple averaging.
(3) Adaptive weighting further improves robustness by capturing dataset-specific field importance, and the penalty term enforces functional correctness by penalizing tools with missing required parameters, which is essential for successful downstream execution.
These results suggest that the components are complementary rather than uniformly additive. 
Their combination in MFTR balances semantic relevance, field reliability, and tool executability.

\begin{table}[t]
  \centering
  \small
  \caption{
    Ablation study of the MFTR framework. 
    "Stand.", "Rewrite", and "Weighting" denote Tool Documentation Standardization, Query Rewriting and Alignment, and Adaptive Weighting, respectively.
    Results are reported in NDCG@10 and the retriever is MiniLM-L6.
    The best performance is highlighted in \textbf{bold}, and the second best is \underline{underlined}.
  }
  \resizebox{\linewidth}{!}
  {
    \begin{tabular}{lccc}
    \toprule
    \textbf{Variant} & \textbf{ToolBench} & \textbf{APIGen} & \textbf{APIBank} \\
    \midrule
    {Full-Doc} & 0.2790 & 0.7126 & 0.6112 \\
    \midrule
        Stand. Only & 0.2941 & 0.6484 & 0.5774 \\
        Rewrite Only & 0.2995 & 0.7063 & 0.6842 \\
        Stand. + Rewrite & 0.2878 & 0.5896 & 0.6622 \\
      \midrule
        Weighting Only & 0.3458 & 0.7186 & 0.5926 \\
      
        Weighting + Stand.  & 0.4231 & 0.8044 & 0.6012 \\
        Weighting + Rewrite  & 0.3253 & 0.8334 & 0.6549 \\
\midrule      {w/o Adaptive Weighting} & 0.5071 & 0.8466 & \underline{0.7138} \\
        {w/o Missing Penalty} & \underline{0.5404} & \textbf{0.8519} & 0.7072 \\
\midrule        {MFTR} & \textbf{0.5412} & \underline{0.8516} & \textbf{0.7237} \\
    \bottomrule
    \end{tabular}%
    }
  \label{tab:exp_ablation}%
\end{table}%

\subsection{Field Analysis}
\label{sec:field_analysis}

\begin{figure}[t]
    \centering
    \includegraphics[width=\linewidth]{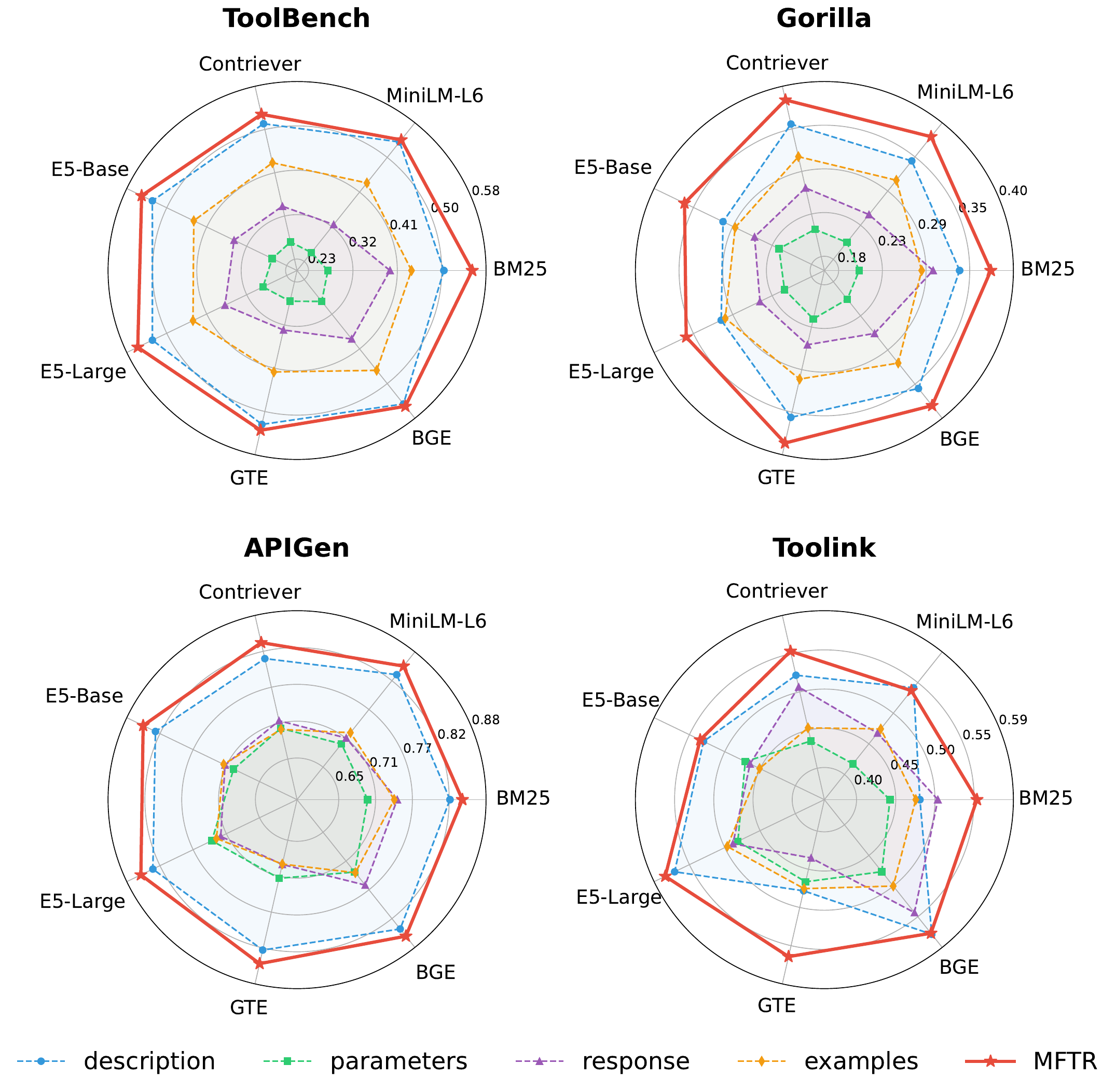}

    \caption{Performance comparison of single-field retrieval with MFTR (Combined). Results are reported in NDCG@10.}
    \label{fig:single_field_radar}
\end{figure}

We analyze single-field retrieval performance using intermediate scores (before adaptive weighting). 
As shown in \cref{fig:single_field_radar}, the \texttt{description} field performs best, aligning with intuition as it concisely summarizes tool functionality. 
The \texttt{examples} field provides concrete usage scenarios, offering the second-most effective information. 
The \texttt{response} field contributes noticeably in certain specific settings, while the \texttt{parameters} field generally exhibits poor performance due to the lack of sufficient contextual information.
Notably, even the strongest single-field variant underperforms MFTR, 
confirming that jointly modeling multiple fields captures richer and  more comprehensive semantic signals.

\subsection{Efficiency Analysis}
\label{sec:efficiency_analysis}

MFTR’s time cost has two phases: an offline stage for one-time documentation standardization and weight training, and an online stage for query rewriting and multi-field retrieval. 
Online, it requires only one LLM call plus lightweight four-field aggregation.
As shown in \cref{fig:time_use} (\cref{appendix:time_use}), MFTR achieves the best precision with moderate latency under the same hardware setup.
Given that agent workflows typically involve multiple LLM generations, MFTR’s marginal latency is an acceptable trade-off for its accuracy gains.

\section{Conclusion}
\label{sec:conclusion}

In this work, we identify that tool retrieval fundamentally differs from ad-hoc text retrieval, as tool utility involves multi-aspect factors beyond semantic similarity.
To address this, we propose Multi-Field Tool Retrieval, a framework that aligns user intent with tool utility via fine-grained multi-field modeling.
MFTR standardizes heterogeneous tool documentation into structured fields, rewrites user queries for structural alignment, and applies adaptive weighting with parameter missing penalties to prioritize functional correctness.
Experiments across five datasets and a large mixed benchmark show that MFTR consistently achieves strong performance, with robust behavior across retrievers and good cross-dataset generalization.
MFTR provides a solid foundation for building more capable and reliable tool-using agents.

\section*{Limitations}

MFTR is designed to improve tool retrieval by aligning user intent with structured tool representations.
While it performs consistently well across benchmarks, several limitations clarify its scope and future directions.
First, MFTR relies on LLMs for documentation standardization and query rewriting. 
While effective, LLM-based reformulation may occasionally introduce hallucinated information absent from the original documentation. 
As noted in \cref{sec:discussion_retrieval_vs_execution}, MFTR targets retrieval effectiveness only; for actual execution, we recommend using original documentation.
Second, while our schema balances functional coverage and generalization, it may not fully capture domain-specific aspects of specialized tools. 
Adaptively extending the schema based on tool characteristics could improve coverage.
Finally, MFTR optimizes retrieval accuracy but does not directly model downstream execution success. 
Although the parameter-missing penalty prioritizes functional constraints, factors like API availability, latency, and error handling remain unaddressed. 
Incorporating execution feedback into retrieval could bridge this gap and enhance end-to-end performance of tool-using agents.

\section*{Acknowledgments}

This work is supported by the Research Project of Quancheng Laboratory, China (Grant No.QCL20250105)


\bibliography{custom}

\appendix

\crefalias{section}{appendixsec}
\crefalias{subsection}{appendixsec}


\begin{table*}[t]
  \small
  \centering
  \caption{Statistics of the datasets. \#TPQ denotes the average number of relevant tools per query. For ToolBench, we removed one of the duplicate queries.}
  \begin{tabular}{l@{\hspace{2pt}}cccl}
    \toprule
    \multicolumn{1}{l}{\textbf{Dataset}} & \textbf{\#Queries} & \textbf{\#Tools} & \textbf{\#TPQ} & \multicolumn{1}{l}{\textbf{Tool Source}}  \\
    \midrule
    ToolBench~\cite{toolbench} & 1099 & 14059 & 2.39 & REST APIs  \\
    APIGen~\cite{apigen} & 1000  & 3605 & 1.33 & REST APIs and Python functions  \\
    APIBank~\cite{apibank} & 101   & 101 & 1.70  & manual Python functions  \\
    Gorilla~\cite{gorilla} (HuggingFace) & 500   & 907  & 1.00 & APIs of HuggingFace models  \\
    Toolink~\cite{toolink} & 497   & 1804 & 2.06 & LLM-generated functions  \\
    \midrule
    Mixed & 3197 & 20476 & 1.77 & All of the above  \\
    \bottomrule
    \end{tabular}
  \label{tab:dataset_statistics}%
\end{table*}%

\section{Experimental Details}
\label{appendix:experimental_setup}

\subsection{Datasets}
\label{appendix:datasets}

\begin{table*}[t]
  \centering
  \small
  \caption{
    Ablation study of the MFTR framework. 
    "Stand.", "Rewrite", and "Weighting" denote Tool Documentation Standardization, Query Rewriting and Alignment, and Adaptive Weighting, respectively.
    Results are reported in NDCG@10.
    The best performance is highlighted in \textbf{bold}, and the second best is \underline{underlined}.
  }
    \begin{tabular}{clccc}
    \toprule
    \textbf{Retriever} & \textbf{Variant} & \textbf{ToolBench} & \textbf{APIGen} & \textbf{APIBank} \\
    \midrule
    \multirow{12}{*}[2pt]{MiniLM-L6} & {Full-Doc} & 0.2790 & 0.7126 & 0.6112 \\
    \cmidrule{2-5}
        & Stand. Only & 0.2941 & 0.6484 & 0.5774 \\
          & Rewrite Only & 0.2995 & 0.7063 & 0.6842 \\
          & Stand. + Rewrite & 0.2878 & 0.5896 & 0.6622 \\
      \cmidrule{2-5}
          & Weighting Only & 0.3458 & 0.7186 & 0.5926 \\
      
          & Weighting + Stand.  & 0.4231 & 0.8044 & 0.6012 \\
          &  Weighting + Rewrite  & 0.3253 & 0.8334 & 0.6549 \\
\cmidrule{2-5}          & {w/o Adaptive Weighting} & 0.5071 & 0.8466 & \underline{0.7138} \\
          & {w/o Missing Penalty} & \underline{0.5404} & \textbf{0.8519} & 0.7072 \\
\cmidrule{2-5}          & {MFTR} & \textbf{0.5412} & \underline{0.8516} & \textbf{0.7237} \\
    \midrule
    \multirow{12}{*}[2pt]{E5-Base} & {Full-Doc} & 0.4722 & 0.7359 & 0.6190 \\
    \cmidrule{2-5}
    & Stand. Only  & 0.4656 & 0.7047 & 0.6549 \\
          & Rewrite Only   & 0.3707 & 0.7297 & 0.6642 \\
          & Stand. + Rewrite    & 0.4695 & 0.7666 & 0.7003 \\
    \cmidrule{2-5}
          & Weighting Only & 0.4088 & 0.7400 & 0.6704 \\
          & Weighting + Stand. & 0.4816 & 0.8149 & 0.6520 \\
          & Weighting + Rewrite & 0.4268 & 0.8345 & 0.5916 \\
\cmidrule{2-5}          & {w/o Adaptive Weighting} & 0.4734 & 0.8495 & \underline{0.7173} \\
          & {w/o Missing Penalty} & \textbf{0.5584} & \underline{0.8513} & 0.7098 \\
\cmidrule{2-5}          & {MFTR} & \underline{0.5520} & \textbf{0.8516} & \textbf{0.7320} \\
    \bottomrule
    \end{tabular}%
  \label{tab:exp_ablation_full}%
\end{table*}%
\begin{table*}[t]
  \small
  \centering
  \caption{
    Ablation study on the background LLM used for tool documentation standardization and query rewriting.
    We compare gpt-4o-mini and Qwen3.5-35B-A3B as the background model.
    We report NDCG@10 (N@10) and Recall@10 (R@10).
  }
  \resizebox{\textwidth}{!}
  {
    \begin{tabular}{c|c||cc|cc|cc|cc|cc}
    \toprule    \multirow{2}{*}[-1.5pt]{\textbf{Retriever}} & \multirow{2}{*}[-1.5pt]{\textbf{LLM}} & \multicolumn{2}{c|}{\textbf{ToolBench}} & \multicolumn{2}{c|}{\textbf{APIGen}} & \multicolumn{2}{c|}{\textbf{APIBank}} & \multicolumn{2}{c|}{\textbf{Gorilla}} & \multicolumn{2}{c}{\textbf{Toolink}} \\
\cmidrule{3-12}          &       & \textbf{N@10} & \textbf{R@10} & \textbf{N@10} & \textbf{R@10} & \textbf{N@10} & \textbf{R@10} & \textbf{N@10} & \textbf{R@10} & \textbf{N@10} & \textbf{R@10} \\
    \toprule
    \multirow{2}[1]{*}{BM25} & gpt-4o-mini & 0.5573 & 0.6270 & 0.8430 & 0.9227 & 0.6708 & 0.6909 & 0.3716 & 0.5620 & 0.5498 & 0.6201 \\ 
          & Qwen3.5 & 0.5297 & 0.6029 & 0.8406 & 0.9202 & 0.6691 & 0.6929 & 0.3714 & 0.5560 & 0.5404 & 0.6027 \\ 
    \midrule
    \multirow{2}[1]{*}{MiniLM-L6} & gpt-4o-mini & 0.5413 & 0.6256 & 0.8516 & 0.9248 & 0.7236 & 0.7862 & 0.3775 & 0.5860 & 0.5334 & 0.6071 \\ 
          & Qwen3.5 & 0.5088 & 0.5928 & 0.8397 & 0.9207 & 0.7095 & 0.7775 & 0.4007 & 0.5980 & 0.5632 & 0.6245 \\ 
    \midrule
    \multirow{2}[1]{*}{Contriever} & gpt-4o-mini & 0.5275 & 0.6103 & 0.8365 & 0.9127 & 0.6816 & 0.7204 & 0.3823 & 0.6000 & 0.5493 & 0.6137 \\ 
          & Qwen3.5 & 0.4961 & 0.5822 & 0.8293 & 0.9067 & 0.6779 & 0.7608 & 0.3765 & 0.5800 & 0.5625 & 0.6396 \\ 
    \midrule
    \multirow{2}[1]{*}{E5-Base} & gpt-4o-mini & 0.5520 & 0.6295 & 0.8516 & 0.9263 & 0.7322 & 0.7631 & 0.3575 & 0.5520 & 0.5315 & 0.5936 \\ 
          & Qwen3.5 & 0.5274 & 0.6073 & 0.8444 & 0.9277 & 0.7089 & 0.7543 & 0.3666 & 0.5500 & 0.5106 & 0.6044 \\ 
    \midrule
    \multirow{2}[1]{*}{E5-Large} & gpt-4o-mini & 0.5606 & 0.6359 & 0.8556 & 0.9302 & 0.7219 & 0.7422 & 0.3551 & 0.5540 & 0.5797 & 0.6510 \\ 
          & Qwen3.5 & 0.5305 & 0.6127 & 0.8474 & 0.9273 & 0.7446 & 0.7790 & 0.3621 & 0.5320 & 0.5162 & 0.6020 \\ 
    \midrule
    \multirow{2}[1]{*}{GTE-Large} & gpt-4o-mini & 0.5348 & 0.6111 & 0.8476 & 0.9272 & 0.6885 & 0.7188 & 0.3849 & 0.5980 & 0.5603 & 0.6131 \\ 
          & Qwen3.5 & 0.5106 & 0.5936 & 0.8302 & 0.9138 & 0.6933 & 0.7259 & 0.3815 & 0.5800 & 0.5323 & 0.6020 \\ 
    \midrule
    \multirow{2}[1]{*}{BGE-Large} & gpt-4o-mini & 0.5544 & 0.6347 & 0.8575 & 0.9308 & 0.7169 & 0.7462 & 0.3796 & 0.5880 & 0.5726 & 0.6490 \\ 
          & Qwen3.5 & 0.5321 & 0.6064 & 0.8489 & 0.9297 & 0.7003 & 0.7435 & 0.3915 & 0.5840 & 0.5944 & 0.6532 \\
    \bottomrule
    \end{tabular}%
  }
  \label{tab:ablation_background_model}%
\end{table*}%

To evaluate the effectiveness of our framework, we conduct experiments on five widely used tool retrieval datasets.
These datasets are originally developed for tool learning and later adapted into retrieval benchmarks by~\citeauthor{shi-etal-2025-toolret}~\cite{shi-etal-2025-toolret}.
Specifically, we include the following datasets:

\begin{itemize}[leftmargin=*]
    \item \textbf{ToolBench}~\cite{toolbench} matches LLM-synthesized instructions with a vast array of real-world REST APIs from RapidAPI.

    \item \textbf{APIGen}~\cite{apigen} improves on ToolBench by systematically cleaning and refining its API documents, and additionally incorporates some Python functions as tools.
    It employs a multi-step verification framework to generate high-quality function calls.

    \item \textbf{APIBank}~\cite{apibank} includes human-authored, multi-turn dialogue queries and manually crafted Python functions as tools.

    \item \textbf{Gorilla}~\cite{gorilla} (HuggingFace subset) 
    focuses on queries that describe desired model capabilities (e.g., text generation), paired with APIs of HuggingFace models.

    \item \textbf{Toolink}~\cite{toolink} pairs task-specific instructions with LLM-generated code snippets produced via few-shot prompting. It reflects the need for retrieving on-the-fly generated functional modules.

\end{itemize}

To provide a clearer overview, \cref{tab:dataset_statistics} summarizes the statistics of these datasets.
To further simulate real-world deployment scenarios, we also construct a \textbf{Mixed} benchmark by merging all five datasets.
This mixed setting comprises diverse queries and a large-scale heterogeneous tool pool, evaluating the robustness and generalization of tool retrieval frameworks.
There are other similar tool retrieval benchmarks with large numbers of tools, such as ToolRet~\cite{shi-etal-2025-toolret}. 
However, we do not adopt ToolRet because it is constructed from a broader and less curated collection, leading to some of its included datasets (e.g., MassTool) containing ambiguous or low-quality tool documentation.
This obscures the true difficulty of the retrieval task and renders retrieval-based evaluation on such datasets less meaningful.

\subsection{Baselines}
\label{appendix:baselines}

We compare MFTR with four baselines:

\begin{itemize}[leftmargin=*]
    \item \textbf{Full-Doc} performs semantic retrieval by directly matching user queries with the complete, raw tool documentation.

    \item \textbf{EasyTool}~\cite{yuan-etal-2025-easytool} leverages LLMs to standardize and enrich tool descriptions and functional guidelines, and performs retrieval based on the augmented documentation.

    \item \textbf{PLUTo}~\cite{pluto} adopts an Edit-and-Ground pipeline to identify and rewrite tool documentation with missing or incomplete information.
    It further employs a Plan-and-Retrieve framework that decomposes queries into sub-queries, retrieves candidate tools with reranking, and uses an LLM to select the optimal tool.

    \item \textbf{OnlineRAG}~\cite{onlinerag} is one of the latest approaches. It introduces a feedback-driven mechanism to continuously refine tool representations. Specifically, it updates tool embeddings in real time according to downstream performance signals, enabling the retriever to better align with actual tool usage behavior over time.
    Since this method needs to update tool embeddings, we do not apply OnlineRAG when using BM25 as the retriever.
        
\end{itemize}

\subsection{Implementation Details}
\label{appendix:implementation_details}

We use gpt-4o-mini as the backbone LLM to perform both tool documentation standardization (\cref{sec:standard_tool_documentation}) and query rewriting (\cref{sec:query_rewriting}). 
We also present the results using Qwen3.5-35B-A3B~\cite{qwen3.5} as the backbone LLM in~\cref{appendix:ablation_llm}.
For LLM generation, we set the temperature to 0, top-p to 1.0, and top-k to 1, with a maximum generation length of 2048 tokens.
For tool-aware query rewriting, we retrieve the top 20 candidate tools using BM25 as in-context reference.
To ensure a robust evaluation and maximize the utility of the datasets, we adopt a 5-fold cross-validation strategy.
Specifically, the dataset is partitioned into five equal subsets, with each subset serving as the test set in rotation, while the remaining four are used for training.
For the Mixed benchmark, the partitioning is performed independently on the merged dataset, rather than aggregating the splits from individual datasets.
During the training phase, we perform negative sampling by selecting the top 64 incorrect tools retrieved by BM25 for each query to serve as hard negatives.
The multi-field relevance computation model (\cref{sec:adaptive_weight}) is trained for 5 epochs with a batch size of 256.
We use the Adam optimizer with a learning rate of 0.1.
Regarding the hyperparameters in our penalty mechanism, we fix the sharpness factor $\alpha=15$.
All experiments were conducted using PyTorch on NVIDIA A800 GPUs with 80GB of memory.

\section{Additional Results and Analysis}
\label{appendix:experiments}

\newcolumntype{L}[1]{>{\raggedright\arraybackslash}m{#1}}

\begin{table*}[t]
  \renewcommand{\arraystretch}{1.1}

  \centering
  \caption{
    Case study on the Gorilla dataset using BM25.
    Compared with Full-Doc, which ranks the golden tool at 24, MFTR leverages multi-field modeling to suppress semantic noise and correctly ranks the tool at position 1.
  }
  \resizebox{\linewidth}{!}
  {
    \begin{tabular}{c c  L{4cm} @{\hspace{-1pt}} L{5cm} @{\hspace{1pt}} ccc}
    \toprule
    \textbf{Query} & \multicolumn{5}{l}{
    
        \begin{minipage}[c]{1.1\linewidth}
            A podcast producer is looking to improve the quality of their audio files by removing background noise. What can they do?
        \end{minipage}
    }
    & \textbf{Final Rank} \\
    [10pt]
    \toprule
    \multirow{1.2}[0]{*}{\textbf{Full-Doc}} &

        \multicolumn{5}{l}{
            \begin{minipage}[c]{1.1\linewidth}
                \texttt{description}: perform speech enhancement (denoising) with a SepFormer model, implemented with SpeechBrain, and pretrained on WHAM! dataset with 16k sampling frequency...{\color{gray}\footnotesize
                \ [Implementation Details: SepFormer model, SpeechBrain framework, WHAM! dataset, 16kHz]
                }
            \end{minipage}
        }                      & \multirow{1.2}[0]{*}{\textbf{Rank: 24}} \\
    [10pt]
    \midrule
    \multirow{12}[0]{*}{\textbf{MFTR}} & \textbf{Field}
          
          & \textbf{Rewrite Query} & \textbf{Tool Documentation} & \textbf{Field Rank} & \textbf{Field Weight} & \multirow{12}[0]{*}{\textbf{Rank: 1}} \\

    \cmidrule{2-6}
    
    & \texttt{description} & 
        Perform speech enhancement on audio files...
        & 
        performs speech enhancement by denoising audio...
        & 2   & 21.4\% &  \\
        
    \cmidrule{2-6}
        
    & \texttt{parameters} & \texttt{audio\_file\_path} & \texttt{path}: The file path to the input audio file. & 1   & 13.3\% &  \\

    \cmidrule{2-6}
    
    & \texttt{response} & An enhanced audio file &   produces an enhanced version of an audio signal.      & 5   & 23.0\% &  \\

    \cmidrule{2-6}
    
    & \texttt{examples} & clean up podcast recordings by removing background noise & clean up a noisy recording of a meeting for transcription & 2  & 42.3\%  &  \\

    \bottomrule
    
    \end{tabular}
  } 
  \label{tab:case_study}%
\end{table*}%

\subsection{Ablation Study}

In this section, we provide detailed results for~\cref{sec:ablation_study}, including two models MiniLM-L6 and E5-Base.
The results are reported in \cref{tab:exp_ablation_full}.
We observe three key findings:

First, LLM-based textual augmentation alone provides limited and unstable gains.
Variants applying Stand. or Rewrite in the Full-Doc paradigm (\textit{Stand. Only}, \textit{Rewrite Only}, \textit{Stand. + Rewrite}) fail to produce consistent improvements and sometimes degrade performance (e.g., APIGen)
This indicates that simply LLM-based augmentation increases the amount of information but introduces semantic noise.

Second, documentation standardization and structural alignment are prerequisites.
The \textit{Weighting + Stand.} variant, which performs retrieval over standardized fields using the original query, steadily outperforms the Full-Doc baseline.
In contrast, the \textit{Weighting + Rewrite} variant fails as it matches structured queries against raw, unaligned documentation.
Similarly, the \textit{Weighting Only} variant, which applies adaptive weights to raw documentation fields, exhibits fluctuating results, showing that adaptive weighting alone cannot compensate for raw documentation noise.
These results confirm that effective multi-field matching relies on a unified schema.
When both standardization and rewriting are enabled, fields can be modeled independently, resulting in substantial performance gains even with simple average aggregation.

Finally, adaptive weighting improves both functional correctness and robustness.
It consistently outperforms simple averaging by capturing varying field importance across datasets.
Regarding the penalty mechanism, while removing it occasionally results in minor fluctuations, the penalty term is crucial for datasets like APIBank.
More importantly, it enforces functional correctness by penalizing tools with missing required parameters, which is essential for successful downstream execution.

\subsection{Ablation Study on Background LLM}
\label{appendix:ablation_llm}

\begin{table*}[t]
  \small
  \centering
  \caption{Distribution of learned field weights.}
    \begin{tabular}{ccccccc}
    \toprule
    \multicolumn{1}{c}{\multirow{2}[0]{*}{\textbf{Field}}} & \multicolumn{3}{c}{\textbf{E5-Base}} & \multicolumn{3}{c}{\textbf{GTE}} \\
    \cmidrule{2-7}
          & \textbf{APIGen} & \textbf{APIBank} & \textbf{Toolink} & \textbf{APIGen} & \textbf{APIBank} & \textbf{Toolink} \\
    \midrule
    \texttt{description} & 26.90\% & 26.40\% & 29.40\% & 38.00\% & 48.00\% & 24.50\% \\
    \texttt{parameters} & 22.10\% & 24.80\% & 9.70\% & 27.80\% & 23.70\% & 24.10\% \\
    \texttt{response} & 24.80\% & 24.30\% & 53.00\% & 19.70\% & 6.30\% & 18.00\% \\
    \texttt{examples} & 26.20\% & 24.60\% & 7.90\% & 14.50\% & 22.00\% & 33.40\% \\
    \bottomrule
    \end{tabular}
  \label{tab:field_weight}%
\end{table*}%

Overall, gpt-4o-mini yields consistently stronger or comparable performance in most retriever--dataset combinations, indicating that higher-quality restructuring and rewriting generally translate into better multi-field matching.
Meanwhile, Qwen3.5-35B-A3B remains competitive and occasionally achieves comparable results, suggesting MFTR is not overly sensitive to the specific background LLM and can benefit from alternative open/efficient models when deployment constraints apply.

We investigate how the choice of the underlying LLM influences MFTR, focusing on two LLM-involved stages: tool documentation standardization and tool-oriented query rewriting.
To this end, we compare gpt-4o-mini with Qwen3.5-35B-A3B~\cite{qwen3.5}, while keeping the retriever backbone, training settings unchanged. 
The results, reported in~\cref{tab:ablation_background_model}, present NDCG@10 and Recall@10 across all five datasets.
Overall, Qwen3.5-35B-A3B and gpt-4o-mini achieve comparable performance in most settings, and both consistently outperform the baseline. 
This observation suggests that MFTR is not particularly sensitive to the specific choice of the background LLM.

\subsection{Case Study}
\label{sec:case_study}

To illustrate the working mechanism of MFTR, we present a case study from the Gorilla dataset using BM25 in \cref{tab:case_study}.
For this query, the Full-Doc baseline ranks the golden tool at position 24, whereas MFTR promotes it to rank 1.
Raw documentation contains technical details (e.g., pretraining datasets) that obscure the core functionality ``speech denoising'', causing a suboptimal ranking.
In contrast, MFTR structures this into fields: \texttt{description} highlights speech enhancement, \texttt{parameters} align with audio inputs, \texttt{response} specifies expected outputs, and \texttt{examples} match the denoising scenario.
By aligning the rewritten query with these granular fields and aggregating field-level signals, MFTR suppresses noise and correctly identifies the target tool.

\subsection{Field Analysis}

We further examine the field weight distributions learned by the adaptive weighting module.
As illustrated in \cref{tab:field_weight}, field importance varies significantly across datasets and retrievers.
In most cases, \texttt{description} accounts for a higher weight, while the remaining fields exhibit greater variability. 
By dynamically adjusting field importance, MFTR effectively accommodates such variability, thereby enhancing retrieval robustness and generalization.

\begin{table*}[t]
  \small
  \centering
  \caption{
    Skill retrieval performance on SRA-Bench datasets.
    N@10 and R@10 denote NDCG@10 and Recall@10, respectively.
    \textbf{Bold} and \underline{underline} indicate the best and second-best results among all methods for each retriever.
    EasyTool uses LLM-augmented skill documentation for retrieval.
    Note that OnlineRAG requires updating tool embeddings, so we do not apply it when using BM25 as the retriever.
  }
  \resizebox{\textwidth}{!}
  {
    \begin{tabular}{c|c||cc|cc|cc|cc|cc}
    \toprule
    \multirow{2}{*}[-1.5pt]{\textbf{Model}} & \multirow{2}{*}[-1.5pt]{\textbf{Method}} & \multicolumn{2}{c|}{\textbf{TheoremQA}} & \multicolumn{2}{c|}{\textbf{LogicBench}} & \multicolumn{2}{c|}{\textbf{ToolQA}} & \multicolumn{2}{c|}{\textbf{ChAMP}} & \multicolumn{2}{c}{\textbf{MedCalcBench}} \\
\cmidrule{3-12}
          &       & \textbf{N@10} & \textbf{R@10} & \textbf{N@10} & \textbf{R@10} & \textbf{N@10} & \textbf{R@10} & \textbf{N@10} & \textbf{R@10} & \textbf{N@10} & \textbf{R@10} \\
    \toprule
    \multirow{5}[1]{*}{BM25}
          & Full-Doc & 0.7479 & \underline{0.8581} & \underline{0.1677} & \underline{0.2684} & \underline{0.6348} & \underline{0.8056} & \underline{0.3641} & \underline{0.4598} & 0.4641 & 0.6318 \\
          & EasyTool & 0.4322 & 0.5234 & 0.0152 & 0.0276 & 0.2844 & 0.4056 & 0.1177 & 0.1446 & 0.4973 & 0.6827 \\
          & PLUTo & \underline{0.7563} & 0.8072 & 0.1049 & 0.1250 & 0.5320 & 0.6021 & 0.2724 & 0.3341 & \underline{0.8201} & \underline{0.8782} \\
          & OnlineRAG & - & - & - & - & - & - & - & - & - & - \\
          & MFTR & \textbf{0.9098} & \textbf{0.9759} & \textbf{0.4066} & \textbf{0.6355} & \textbf{0.8214} & \textbf{0.9266} & \textbf{0.4640} & \textbf{0.6233} & \textbf{0.9760} & \textbf{1.0000} \\
    \midrule
    \multirow{5}[1]{*}{E5-Base}
          & Full-Doc & 0.7457 & 0.8474 & 0.0883 & 0.1526 & 0.5016 & 0.6741 & 0.2589 & 0.3957 & 0.4882 & 0.6200 \\
          & EasyTool & 0.4996 & 0.5997 & 0.0110 & 0.0250 & 0.2974 & 0.3881 & 0.1605 & 0.2387 & 0.3904 & 0.5545 \\
          & PLUTo & \underline{0.8303} & \underline{0.8755} & \underline{0.1970} & \underline{0.2211} & \underline{0.6416} & \underline{0.7266} & \underline{0.4024} & \underline{0.4996} & \underline{0.7259} & \underline{0.7373} \\
          & OnlineRAG & 0.7456 & 0.8460 & 0.0878 & 0.1526 & 0.5012 & 0.6734 & 0.2598 & 0.4002 & 0.4878 & 0.6200 \\
          & MFTR & \textbf{0.8941} & \textbf{0.9665} & \textbf{0.4247} & \textbf{0.6737} & \textbf{0.8210} & \textbf{0.9336} & \textbf{0.4541} & \textbf{0.6125} & \textbf{0.9830} & \textbf{1.0000} \\
    \midrule
    \multirow{5}[1]{*}{E5-Large}
          & Full-Doc & 0.7631 & 0.8635 & 0.1500 & 0.2368 & 0.5337 & 0.7063 & 0.2794 & 0.3813 & 0.5280 & 0.6727 \\
          & EasyTool & 0.5232 & 0.6212 & 0.0179 & 0.0342 & 0.2673 & 0.3587 & 0.1718 & 0.2466 & 0.3282 & 0.4164 \\
          & PLUTo & \underline{0.8531} & \underline{0.8929} & \underline{0.2690} & \underline{0.3145} & \underline{0.6611} & \underline{0.7434} & \underline{0.4047} & \underline{0.4830} & \underline{0.7745} & \underline{0.7909} \\
          & OnlineRAG & 0.7637 & 0.8635 & 0.1493 & 0.2368 & 0.5326 & 0.7035 & 0.2777 & 0.3768 & 0.5290 & 0.6727 \\
          & MFTR & \textbf{0.8922} & \textbf{0.9625} & \textbf{0.4527} & \textbf{0.7079} & \textbf{0.8404} & \textbf{0.9420} & \textbf{0.4220} & \textbf{0.5358} & \textbf{0.9877} & \textbf{1.0000} \\
    \midrule
    \multirow{5}[1]{*}{GTE-Large}
          & Full-Doc & 0.8323 & 0.9291 & 0.3711 & 0.6461 & 0.6140 & 0.7881 & 0.3986 & 0.5435 & 0.8482 & 0.9355 \\
          & EasyTool & 0.5779 & 0.6867 & 0.1077 & 0.2355 & 0.2785 & 0.4322 & 0.2443 & 0.3701 & 0.7256 & 0.8173 \\
          & PLUTo & \underline{0.8819} & \underline{0.9491} & \textbf{0.6084} & \textbf{0.7434} & \underline{0.7032} & \underline{0.8154} & \underline{0.4844} & \underline{0.5999} & \underline{0.9475} & \underline{0.9700} \\
          & OnlineRAG & 0.8319 & 0.9291 & 0.3701 & 0.6461 & 0.6136 & 0.7874 & 0.3982 & 0.5420 & 0.8481 & 0.9355 \\
          & MFTR & \textbf{0.9031} & \textbf{0.9786} & \underline{0.4113} & \underline{0.6579} & \textbf{0.8455} & \textbf{0.9448} & \textbf{0.5123} & \textbf{0.6556} & \textbf{0.9791} & \textbf{1.0000} \\
    \midrule
    \multirow{5}[1]{*}{BGE-Large}
          & Full-Doc & 0.7694 & 0.8701 & 0.1013 & 0.1776 & 0.5297 & 0.7182 & 0.2838 & 0.4320 & 0.4988 & 0.6164 \\
          & EasyTool & 0.5401 & 0.6493 & 0.0264 & 0.0526 & 0.3742 & 0.4776 & 0.1879 & 0.2688 & 0.3882 & 0.5027 \\
          & PLUTo & \underline{0.8510} & \underline{0.9050} & \underline{0.2213} & \underline{0.2553} & \underline{0.6534} & \underline{0.7441} & \underline{0.3818} & \underline{0.4974} & \underline{0.7562} & \underline{0.7682} \\
          & OnlineRAG & 0.7697 & 0.8701 & 0.1006 & 0.1763 & 0.5309 & 0.7182 & 0.2852 & 0.4309 & 0.4988 & 0.6164 \\
          & MFTR & \textbf{0.8956} & \textbf{0.9665} & \textbf{0.3593} & \textbf{0.6184} & \textbf{0.8698} & \textbf{0.9755} & \textbf{0.5033} & \textbf{0.6590} & \textbf{0.9846} & \textbf{1.0000} \\
    \bottomrule
    \end{tabular}%
  }
  \label{tab:skill_overall}%
\end{table*}%

\subsection{Efficiency Analysis}
\label{appendix:time_use}

\begin{figure}[t]
    \centering
    \includegraphics[width=\linewidth]{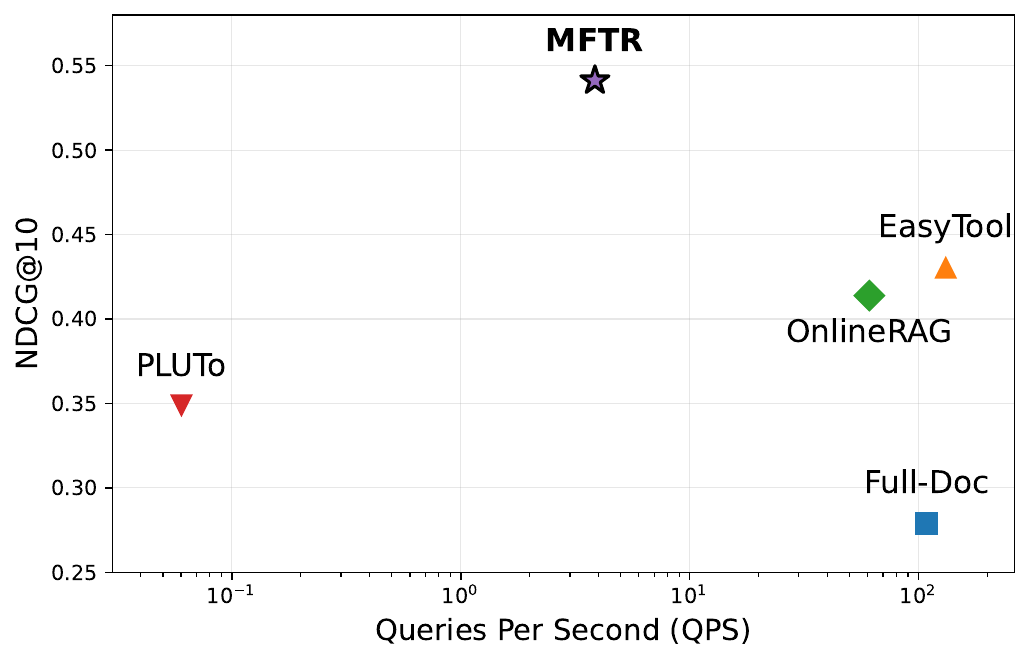}

    \caption{Comparison of Queries Per Second (QPS) and NDCG@10 on the ToolBench dataset using the retriever MiniLM-L6. The x-axis is on a logarithmic scale.}
    \label{fig:time_use}
\end{figure}

MFTR’s time cost consists of an offline stage (one‑time document standardization and weight training) and an online stage (query rewriting plus multi‑field retrieval). 
In the online stage, MFTR makes only a single LLM call for query rewriting, followed by lightweight aggregation over four fields.
As shown in \cref{fig:time_use}, in the same hardware environment, MFTR achieves the best NDCG@10 while maintaining a moderate QPS. 
Given that agent workflows typically involve multiple LLM generations, MFTR’s marginal latency is an acceptable trade‑off for its accuracy gains.

\section{MFTR on Skill Retrieval}
\label{appendix:skill_retrieval}

To extend the applicability of MFTR beyond tool repositories, we further study its effectiveness on skill retrieval for agents.
Compared with tool retrieval, skill retrieval focuses on selecting reusable agent skills (i.e., procedural knowledge or solution patterns) that can guide downstream reasoning and execution.
We evaluate MFTR and tool retrieval baselines on SRA-Bench~\cite{su2026skillretrievalaugmentationagentic}, which is a comprehensive benchmark suite designed for skill retrieval and augmentation for agentic problem solving.
We use five sub-datasets covering diverse domains: \textbf{TheoremQA} (theorem-based reasoning), \textbf{LogicBench} (formal logic question answering), \textbf{ToolQA} (tool-interleaved question answering), \textbf{ChAMP} (competitive algorithmic problem solving) and \textbf{MedCalcBench} (medical calculation tasks).
Across these sub-datasets, each query is associated with a pool of candidate skills described in natural language, and the goal is to rank skills that best match the query intent.

\paragraph{Adapting MFTR to skill retrieval.}
Since agent skills typically do not possess explicit parameter specifications, we adapt MFTR by retaining three fields: \texttt{description}, \texttt{response}, and \texttt{examples}.
We standardize skill documentation and rewrite queries following the same methodology described in \cref{sec:standard_tool_documentation,sec:query_rewriting}, and aggregate the three-field relevance scores using adaptive weighting as in \cref{sec:adaptive_weight}.
All experimental settings remain consistent with the tool retrieval experiments described in \cref{appendix:implementation_details}, except that we use Qwen3.5-35B-A3B~\cite{qwen3.5} as the backbone LLM for documentation standardization and query rewriting.

\paragraph{Results and analysis.}
The results are shown in~\cref{tab:skill_overall}.
MFTR consistently achieves the best or second-best performance in most settings, demonstrating strong generalizability to the skill retrieval task.
Notably, MFTR achieves near-perfect performance on MedCalcBench across all retrievers (NDCG@10 up to 0.9877 and Recall@10 reaching 1.0000), indicating that multi-field relevance modeling effectively captures the precise functional matching required in specialized domains.
On TheoremQA and ToolQA, MFTR substantially outperforms all baselines, with improvements consistent with the patterns observed in tool retrieval experiments (\cref{sec:main_results}).
PLUTo, which employs an LLM-based tool selection mechanism, occasionally achieves competitive results, suggesting that certain skill retrieval scenarios may benefit from more aggressive query--skill interaction modeling.
Furthermore, MFTR with BM25 often matches or surpasses larger embedding models (e.g., GTE-Large), reinforcing the conclusion gains primarily stem from multi-field relevance modeling rather than model capacity.
These results support our conclusion that MFTR is not limited to tool retrieval. 
As a general multi-field relevance modeling framework, it also transfers effectively to skill retrieval, providing strong and consistent improvements under diverse datasets and retriever architectures.

\section{Tool Documentation Examples}
\label{appendix:tool_document_case}

Tool documentation collected from different sources often varies substantially in structure, terminology, and granularity.
To make this heterogeneity concrete, we present representative examples from the five evaluation datasets.

\noindent These examples highlight the severe cross-source heterogeneity of tool documentation. 
Different repositories adopt different schemas and naming conventions (e.g. explicit \texttt{parameters} objects in APIGen, and list-style parameter specifications plus \texttt{responses} in APIBank), and may expose distinct auxiliary fields (e.g., \texttt{domain}, \texttt{framework}, \texttt{api\_call} in Gorilla). 
The granularity is also inconsistent. 
Some entries provide detailed input/output specifications, while others only contain high-level descriptions or code snippets (e.g., \texttt{func\_description} in Toolink), and some even miss key functional details (e.g., \texttt{api\_arguments}: \texttt{N/A} in Gorilla). 
Such discrepancies make it difficult to directly align tools from different sources into a shared semantic space, and motivate the need for documentation standardization in MFTR.

\DefineVerbatimEnvironment{ToolExample}{Verbatim}{
  breaklines=true,
  breakanywhere=true,
  breakautoindent=true,
  breakindent=1.3em,
  breaksymbolleft={},
  breaksymbolright={},
  frame=single,
  fontsize=\small,
}

\textbf{ToolBench.} Original documentation:
\begin{ToolExample}
{
  "description": "# Returns an instance of a gspread.cell.Cell located at row and col column. \nworksheet.cell(row, col, value_render_option='FORMATTED_VALUE')\n\nExamples:\n>>> worksheet.cell(1, 1)\n<Cell R1C1 \"I'm cell A1\">\n",
  "name": "worksheet.cell"
}
\end{ToolExample}

\noindent Standardized documentation:
\begin{ToolExample}
{
  "description": "Retrieve a specific cell from a spreadsheet by its row and column, returning its contents and properties for inspection or manipulation.",
  "parameters": {
    "row": {
      "type": "number",
      "description": "The row number of the cell to retrieve, starting from 1 for the first row.",
      "required": true,
      "example": 1
    },
    "col": {
      "type": "number",
      "description": "The column number of the cell to retrieve, starting from 1 for the first column.",
      "required": true,
      "example": 1
    },
    "value_render_option": {
      "type": "string",
      "description": "Determines how the cell's value should be returned. Common options include 'FORMATTED_VALUE' to return the displayed value, or other rendering options supported by the API.",
      "required": false,
      "default": "FORMATTED_VALUE",
      "example": "FORMATTED_VALUE"
    }
  },
  "response": "The tool returns an instance of a `gspread.cell.Cell`, which represents a cell in a Google Sheets worksheet located at the specified row and column. The response encapsulates information about the cell at the given position, such as its value.\n\nThe components of the response include:\n\n1. **Cell Reference**: The specific row and column of the cell, which determines its position in the worksheet. This is implicitly defined by the `row` and `col` parameters passed to the function.\n\n2. **Cell Value**: The value contained in the specified cell. By default, this may be the formatted value (`FORMATTED\VALUE`), which is the human-readable representation of the cell's contents as displayed in the worksheet.\n\nThe exact structure of the response would typically be an object or reference to a `Cell` instance, with properties such as:\n- **Row and Column Information**: Indicates the cell\u2019s position.\n- **Value or Content**: The data held within the cell at the specified location.\n\nThe response encapsulates the cell's content, which can vary based on the specific contents and formatting of the cell in the worksheet.\n\n",
  "examples": [
    {
      "user_intent": "Get the value from the cell at row 2, column 3 in the worksheet.",
      "tool_call": {
        "name": "worksheet.cell",
        "arguments": {
          "row": 2,
          "col": 3
        }
      }
    },
    {
      "user_intent": "Retrieve the value in the first row and first column of the worksheet.",
      "tool_call": {
        "name": "worksheet.cell",
        "arguments": {
          "row": 1,
          "col": 1
        }
      }
    },
    {
      "user_intent": "I want to fetch the formatted value of the cell located at row 4, column 5.",
      "tool_call": {
        "name": "worksheet.cell",
        "arguments": {
          "row": 4,
          "col": 5,
          "value_render_option": "FORMATTED_VALUE"
        }
      }
    }
  ]
}
\end{ToolExample}

\textbf{APIGen.} Original documentation:
\begin{ToolExample}
{
  "name": "live_giveaways_by_type",
  "description": "Retrieve live giveaways from the GamerPower API based on the specified type.",
  "parameters": {
    "type": {
      "description": "The type of giveaways to retrieve (e.g., game, loot, beta).",
      "type": "str",
      "default": "game"
    }
  }
}
\end{ToolExample}

\noindent Standardized documentation:
\begin{ToolExample}
{
  "description": "Retrieve ongoing giveaways from the GamerPower API filtered by a specific type, such as games, loot, or beta access, to quickly find relevant promotions.",
  "parameters": {
    "type": {
      "type": "string",
      "description": "The type of giveaways to retrieve. This can include options such as 'game', 'loot', or 'beta'.",
      "required": true,
      "default": "game",
      "example": "game"
    }
  },
  "response": "The tool returns a collection of live giveaways filtered by the specified type. Each giveaway in the response typically includes key information such as the giveaway title, description, platform, eligibility criteria, start and end dates, and a link or instructions to participate. The response may also include metadata about the number of active giveaways returned or pagination details if the dataset is large. The primary purpose of the response is to provide a structured list of ongoing giveaways that match the requested type.",
  "examples": [
    {
      "user_intent": "Show me all the current game giveaways that are live right now.",
      "tool_call": {
        "name": "live_giveaways_by_type",
        "arguments": {
          "type": "game"
        }
      }
    },
    {
      "user_intent": "I want to find any live beta access giveaways available for new games.",
      "tool_call": {
        "name": "live_giveaways_by_type",
        "arguments": {
          "type": "beta"
        }
      }
    },
    {
      "user_intent": "Are there any loot giveaways happening at the moment?",
      "tool_call": {
        "name": "live_giveaways_by_type",
        "arguments": {
          "type": "loot"
        }
      }
    }
  ]
}
\end{ToolExample}

\textbf{Gorilla.} Original documentation:
\begin{ToolExample}
{
  "domain": "Natural Language Processing Feature Extraction",
  "framework": "Hugging Face Transformers",
  "functionality": "Feature Extraction",
  "api_call": "AutoModel.from_pretrained('YituTech/conv-bert-base')",
  "api_arguments": "N/A",
  "python_environment_requirements": "transformers",
  "example_code": "N/A",
  "performance": {
    "dataset": "N/A",
    "accuracy": "N/A"
  },
  "description": "A pre-trained ConvBERT model for feature extraction provided by YituTech, based on the Hugging Face Transformers library.",
  "name": "YituTech/conv-bert-base"
}
\end{ToolExample}

\noindent Standardized documentation:
\begin{ToolExample}
{
  "description": "A pre-trained ConvBERT model from YituTech, available via Hugging Face Transformers, designed to extract rich text features for natural language processing tasks.",
  "parameters": {},
  "response": "The tool does not explicitly define a response format in the provided documentation. Based on its described functionality—feature extraction using a pre-trained ConvBERT model—the typical response would be a structured numerical representation of input text suitable for downstream NLP tasks. Specifically:\n\n- **Type of response**: A multi-dimensional numerical tensor or array.  \n- **Components**:  \n  - **Batch dimension (optional)**: If multiple text inputs are processed at once, the first dimension represents the number of inputs.  \n  - **Sequence dimension**: Corresponds to the number of tokens in each input text, determined by tokenization.  \n  - **Feature/hidden dimension**: Each token is represented as a dense vector encoding its semantic and contextual information, typically of fixed size determined by the model’s hidden layer size.  \n- **Semantic meaning**: Each vector captures the contextualized embedding of a token or, if pooled, of the entire input text. These embeddings can be used for downstream tasks such as classification, similarity computation, clustering, or other NLP analyses.  \n- **Additional notes**: The tool itself does not return labels, predictions, or accuracy metrics—it only produces raw feature embeddings. Post-processing (e.g., pooling or aggregation) may be applied depending on the application.  \n\nIn summary, the response is a tensor of numerical embeddings representing the semantic and contextual features of the input text.",
  "examples": [
    {
      "user_intent": "I want to extract embeddings from a batch of customer reviews using a pre-trained ConvBERT model.",
      "tool_call": {
        "name": "YituTech/conv-bert-base",
        "arguments": {
          "api_call": "AutoModel.from_pretrained('YituTech/conv-bert-base')",
          "input_texts": [
            "The product quality is excellent and delivery was fast.",
            "Customer service was unhelpful and rude."
          ]
        }
      }
    },
    {
      "user_intent": "Generate feature vectors for sentences to use in semantic similarity comparisons.",
      "tool_call": {
        "name": "YituTech/conv-bert-base",
        "arguments": {
          "api_call": "AutoModel.from_pretrained('YituTech/conv-bert-base')",
          "input_texts": [
            "I love hiking in the mountains.",
            "Walking through nature trails is very relaxing."
          ]
        }
      }
    },
    {
      "user_intent": "Use ConvBERT embeddings to classify news articles into categories.",
      "tool_call": {
        "name": "YituTech/conv-bert-base",
        "arguments": {
          "api_call": "AutoModel.from_pretrained('YituTech/conv-bert-base')",
          "input_texts": [
            "The government passed a new economic stimulus package today.",
            "The local team won the championship in a thrilling final match."
          ]
        }
      }
    }
  ]
}
\end{ToolExample}

\textbf{APIBank.} Original documentation:
\begin{ToolExample}
{
  "category": "External impact - database operation",
  "parameters": [
    {
      "name": "username",
      "in": "query",
      "schema": {
        "type": "str",
        "default": null
      },
      "description": "The username of the user."
    },
    {
      "name": "password",
      "in": "query",
      "schema": {
        "type": "str",
        "default": null
      },
      "description": "The password of the user."
    }
  ],
  "description": "\"Get user token API that takes two parameters - 'username' and 'password'. The 'username' parameter is used to identify the user and the 'password' parameter is used to authenticate the user's identity.\"",
  "url": "apis/get_user_token.py",
  "method": "GET",
  "responses": {
    "token": {
      "type": "str",
      "description": "The token generated upon successful authentication."
    }
  },
  "name": "GetUserToken"
}
\end{ToolExample}

\noindent Standardized documentation:
\begin{ToolExample}
{
  "description": "Authenticate a user by verifying their username and password, and generate a secure token for accessing protected resources.",
  "parameters": {
    "username": {
      "type": "string",
      "description": "The username of the user used to identify them for authentication.",
      "required": true,
      "example": "john_doe"
    },
    "password": {
      "type": "string",
      "description": "The password of the user used to authenticate their identity.",
      "required": true,
      "example": "P@ssw0rd123"
    }
  },
  "response": "The tool returns a response containing a single component:\n\n- **token**: A string representing the authentication token generated when the provided username and password are successfully validated. This token serves as proof of authentication and can be used for subsequent authorized requests.  \n\nThe response structure is a simple object with one key (`token`), and its value is always expected to be a string.",
  "examples": [
    {
      "user_intent": "I want to log in as Alice to get my access token for the app.",
      "tool_call": {
        "name": "GetUserToken",
        "arguments": {
          "username": "Alice",
          "password": "AliceSecure123!"
        }
      }
    },
    {
      "user_intent": "Retrieve the authentication token for my admin account.",
      "tool_call": {
        "name": "GetUserToken",
        "arguments": {
          "username": "admin_user",
          "password": "AdminPass!2025"
        }
      }
    },
    {
      "user_intent": "Get a user token for testing purposes with the test account.",
      "tool_call": {
        "name": "GetUserToken",
        "arguments": {
          "username": "test_account",
          "password": "Test1234"
        }
      }
    }
  ]
}
\end{ToolExample}

\textbf{Toolink.} Original documentation:
\begin{ToolExample}
{
  "func_description": "```python\ndef multiply_fractions(fraction_1, fraction_2):\n    ...\n    return (numerator, denominator)\n```",
  "name": "multiply_fractions",
  "description": "This method takes two arguments, 'fraction_1' and 'fraction_2', where each fraction is a tuple of two integers representing the numerator and denominator. It returns the result of multiplying the two fractions."
}
\end{ToolExample}

\noindent Standardized documentation:
\begin{ToolExample}
{
  "description": "The \"multiply_fractions\" tool multiplies two fractions, represented as tuples of integers (numerator, denominator), and returns the resulting fraction in tuple form.",
  "parameters": {
    "fraction_1": {
      "type": "tuple",
      "description": "A tuple of two integers representing the numerator and denominator of the first fraction to be multiplied.",
      "required": true,
      "example": [
        1,
        2
      ]
    },
    "fraction_2": {
      "type": "tuple",
      "description": "A tuple of two integers representing the numerator and denominator of the second fraction to be multiplied.",
      "required": true,
      "example": [
        3,
        4
      ]
    }
  },
  "response": "The tool returns a tuple containing two integers: the numerator and denominator of the product of the two input fractions. The response represents the result of multiplying the two fractions provided as input arguments, where the numerator is the result of multiplying the numerators of the input fractions, and the denominator is the result of multiplying the denominators of the input fractions.",
  "examples": [
    {
      "user_intent": "Multiply 3/4 by 2/5 and get the result.",
      "tool_call": {
        "name": "multiply_fractions",
        "arguments": {
          "fraction_1": [
            3,
            4
          ],
          "fraction_2": [
            2,
            5
          ]
        }
      }
    },
    {
      "user_intent": "I want to find the product of 7/8 and 1/3.",
      "tool_call": {
        "name": "multiply_fractions",
        "arguments": {
          "fraction_1": [
            7,
            8
          ],
          "fraction_2": [
            1,
            3
          ]
        }
      }
    },
    {
      "user_intent": "Calculate the multiplication of 5/6 and 9/10 for my math homework.",
      "tool_call": {
        "name": "multiply_fractions",
        "arguments": {
          "fraction_1": [
            5,
            6
          ],
          "fraction_2": [
            9,
            10
          ]
        }
      }
    }
  ]
}
\end{ToolExample}

\section{Prompt Template}

\label{appendix:prompt}

The following are the prompts used in the tool documentation standardization and query rewriting processes.

\begin{figure*}[!htb]
\begin{tcolorbox}[colback=lightgray!20,colframe=darkgray!80,title=Prompt 1: Standardize tool's \texttt{description} field]
\small
\setlength{\baselineskip}{9.5pt}
system:

You are an expert tool-documentation generator for Tool-Augmented LLM systems.

Given a raw tool documentation in JSON (which may be noisy, incomplete, poorly written, or overly literal), your task is to produce a high-quality, embedding-friendly natural-language description of the tool.

The generated description must describe clearly what the tool does and why it is used.

Additional requirements:

1. Accurately reflect the semantics of the original documentation (no hallucinated functionality).

2. Focus on the essential information needed for retrieval.

3. Omit irrelevant implementation details, default values, or verbose JSON structures.

4. Produce a single concise paragraph (no JSON, no markdown, no headings).

Below are examples showing the expected transformation.

Input tool documentation:

\{

\ \ \ \  "name": "search\_items",
  
\ \ \ \ "description": "Tool for searching items in a database.",
  
\ \ \ \  "parameters": \{
  
\ \ \ \ \ \ \ \       "type": "object",
      
\ \ \ \ \ \ \ \       "properties": \{
      
\ \ \ \ \ \ \ \  \ \ \ \          "query": \{ "type": "string", "description": "text to search" \},
          
\ \ \ \ \ \ \ \  \ \ \ \          "limit": \{ "type": "number", "description": "max results" \}
          
\ \ \ \ \ \ \ \       \}
      
\ \ \ \  \}
  
\}

Target output description:

Search a database and retrieves items that match a given text query.

\vspace{3mm}

user:

You are given the following raw tool JSON.

Generate a concise, high-quality natural-language description that accurately describes what the tool does and why it would be used.

Guidelines:

+ Do not repeat unnecessary JSON details.

+ Do not add features that are not present in the JSON.

+ Output only the final description paragraph.

Raw Tool JSON:

\{origin\_doc\}

Please output the improved description.
\end{tcolorbox}

\end{figure*}

\begin{figure*}[!t]
    
\begin{tcolorbox}[colback=lightgray!20,colframe=darkgray!80,title=Prompt 2: Standardize tool's \texttt{parameters} field]
\small
system:

You are an expert parameter-documentation generator for Tool-Augmented LLM systems.

Given a raw tool documentation in JSON (which may be noisy, incomplete, poorly written, or overly literal), your task is to rewrite the `parameters` field into a clean, accurate, high-quality details that preserves all essential parameter information.

For each parameter, clearly describe:

- parameter name

- data type

- description and purpose of the parameter

- whether the parameter is required

- any default value (must retain if present)

- example input values (construct a simple example consistent with the type and description)

- other initial information

Please rewrite parameters into the following structure: \{

\ \ \ \   "<parameter\_name>": \{

\ \ \ \ \ \ \ \       "type": "<data type>",

\ \ \ \ \ \ \ \       "description": "<clear natural-language description of what the parameter means and how it is used>",

\ \ \ \ \ \ \ \       "required": <true or false>,

\ \ \ \ \ \ \ \       "default": "<default value if present, otherwise omit the field>",

\ \ \ \ \ \ \ \       "example": "<a simple example input value consistent with the type>"

\ \ \ \   \}

\}

Your rewritten output must satisfy the following requirements:

1. The final output should contain ONLY the improved valid JSON object for the parameters. No comments, no markdown, no explanations.

1. Do not hallucinate new parameters or remove existing ones.

2. Do not omit any explicit details in the original JSON (e.g., default, allowed options, constraints).

4. The resulting text should be optimized for semantic retrieval and tool-use reasoning.

5. Avoid exposing schema-level fields such as "properties", "type": "object", etc.

6. The "required" field should be determined from the raw JSON. If no such information exists, mark all parameters as required (`required: true`).

Your output should read like a natural-language explanation of all inputs required by the tool, suitable for use in an embedding model.

Below are examples showing the expected rewritten.

Input tool documentation: \{

\ \ \ \   "name": "search\_items",
  
\ \ \ \   "description": "Tool for searching items in a database.",
  
\ \ \ \   "parameters": \{
  
 \ \ \ \ \ \ \ \      "type": "object",
      
 \ \ \ \ \ \ \ \      "properties": \{
      
 \ \ \ \ \ \ \ \ \ \ \ \          "query": \{ "type": "string", "description": "text to search" \},
          
\ \ \ \ \ \ \ \ \ \ \ \           "limit": \{ "type": "number", "description": "max results" \}
          
\ \ \ \ \ \ \ \      \}
      
\ \ \ \   \}
  
\}

Target rewritten parameters in JSON format: \{

\ \ \ \   "query": \{
  
\ \ \ \ \ \ \ \       "type": "string",
      
\ \ \ \ \ \ \ \       "description": "A text string used as the search keyword when querying the database.",
      
\ \ \ \ \ \ \ \       "required": true,
      
\ \ \ \ \ \ \ \       "example": "apple"
      
\ \ \ \     \},
    
\ \ \ \     "limit": \{
    
 \ \ \ \ \ \ \ \      "type": "number",
      
\ \ \ \ \ \ \ \       "description": "The maximum number of results to return from the database search.",
      
\ \ \ \ \ \ \ \       "required": true,
      
\ \ \ \ \ \ \ \       "example": 10
      
\ \ \ \   \}
  
\}

\vspace{1mm}

user:

You are given the following raw tool JSON.

Rewrite ONLY the 'parameters' section into the improved JSON format described in the instructions.

Output only the resulting JSON object and nothing else.

Raw Tool JSON:

\{origin\_doc\}
\end{tcolorbox}

\end{figure*}

\begin{figure*}
    
\begin{tcolorbox}[colback=lightgray!20,colframe=darkgray!80,title=Prompt 3: Standardize \texttt{response} field]
\small
system:

You are an expert tool-documentation generator for Tool-Augmented LLM systems.

Given a raw tool documentation in JSON, your task is to describe the expected content of the tool’s response.

Your goal is to explain what the tool’s response contains and what each part means in principle. Focus on explaining the semantic content and role of each component in the response.

Do not generate a concrete instance of the response.

If the tool does not return a response, simply state that there is no response.

If the documentation explicitly defines the response structure or fields, you should describe them clearly and in detail.

If the documentation does not fully specify the response, you should infer and describe a plausible response schema and content types that are consistent with the documented functionality.

Additional requirements:

1. Do not fabricate specific values, examples, or instance-level data.

2. Prefer abstract descriptions of structure, fields, and semantic roles over concrete outputs.

1. Strictly adhere to the semantics and constraints described in the documentation.

4. Use precise, neutral language suitable for downstream modeling or retrieval.

Below are examples showing the expected transformation.

Input tool documentation:

\{

\ \ \ \   "name": "search\_items",

\ \ \ \   "description": "Tool for searching items in a database.",

\ \ \ \   "parameters": \{

\ \ \ \ \ \ \ \       "type": "object",

\ \ \ \ \ \ \ \       "properties": \{

 \ \ \ \ \ \ \ \ \ \ \ \          "query": \{ "type": "string", "description": "text to search" \},
 
 \ \ \ \ \ \ \ \ \ \ \ \          "limit": \{ "type": "number", "description": "max results" \}
 
\ \ \ \ \ \ \ \       \}

\ \ \ \  \}

\}

Target output description:

The tool returns a collection of database items that satisfy the given text query. Each item in the response represents one search result, with the total number of returned items constrained by the specified maximum result limit.

\vspace{3mm}

user:

You are given the following raw tool JSON.

If the response format is defined, describe its structure and the meaning of each component.

If the response format is not explicitly defined, describe a reasonable and typical form of response the tool would produce, including the kinds of information it would contain.

Guidelines:

+ Do not generate example values or mock data.

+ Output only the final response description.

Raw Tool JSON:

\{origin\_doc\}

Please output the response.
\end{tcolorbox}

\end{figure*}

\begin{figure*}
    
\begin{tcolorbox}[colback=lightgray!20,colframe=darkgray!80,title=Prompt 4: Standardize tool's \texttt{examples} field]
\small
system:

You are an expert scenario-generator for Tool-Augmented LLM systems.

Given a raw tool documentation in JSON, your task is to generate a set of diverse, high-quality, and realistic "Example Uses" for this tool. 

These examples should bridge the gap between a user's natural language intent and the formal tool execution. Each example must consist of:

- user\_intent: A natural language query or request that a user would likely say.

- tool\_call: A precise JSON-formatted call to the tool with realistic argument values.

Guidelines for generation:

1. Diversity: Generate 1-3 distinct examples covering different common use cases (e.g., simple queries, edge cases, or multi-parameter queries).

2. Realism: Use realistic and contextually appropriate values for the parameters, not just placeholders like "string" or "123".

3. Consistency: Ensure all parameters used in the `tool\_call` exist in the raw tool documentation.

4. Accuracy: The `query` should logically lead to the provided `tool\_call`.

5. Format: Output ONLY a valid JSON list of objects. No markdown, no explanations.

Below is an example showing the expected transformation.

Input tool documentation:

\{

\ \ \ \   "name": "search\_items",
  
\ \ \ \   "description": "Tool for searching items in a database.",
  
\ \ \ \   "parameters": \{
  
\ \ \ \ \ \ \ \       "type": "object",
      
\ \ \ \ \ \ \ \       "properties": \{
      
\ \ \ \ \ \ \ \ \ \ \ \           "query": \{ "type": "string", "description": "text to search" \},
          
\ \ \ \ \ \ \ \ \ \ \ \           "limit": \{ "type": "number", "description": "max results" \}
          
\ \ \ \ \ \ \ \       \}
      
\ \ \ \   \}
  
\}

Target output example\_use in JSON format:

[

\ \ \ \   \{

\ \ \ \ \ \ \ \       "user\_intent": "Find some laptops available in the inventory.",

\ \ \ \ \ \ \ \       "tool\_call": \{ "name": "search\_items", "arguments": \{ "query": "laptop", "limit": 5 \} \}

\ \ \ \   \},

 \ \ \ \  \{
 
\ \ \ \ \ \ \ \       "user\_intent": "I need to see the top 10 results for organic coffee beans.",

\ \ \ \ \ \ \ \       "tool\_call": \{ "name": "search\_items", "arguments": \{ "query": "organic coffee beans", "limit": 10 \} \}

 \ \ \ \  \},
 
\ \ \ \   \{

 \ \ \ \ \ \ \ \      "user\_intent": "Search the database for 'ergonomic chairs'.",
 
\ \ \ \ \ \ \ \       "tool\_call": \{ "name": "search\_items", "arguments": \{ "query": "ergonomic chairs" \} \}

\ \ \ \   \}

]

\vspace{3mm}

user:

You are given the following raw tool JSON.

Generate a JSON list of 1-3 high-quality example uses (user\_intent and tool\_call) that accurately demonstrate how to use the tool.

Guidelines:

+ Provide diverse scenarios.

+ Use realistic parameter values.

+ Output ONLY the final JSON list.

Raw Tool JSON:

\{origin\_doc\}

Please output the improved example\_use list.
\end{tcolorbox}

\end{figure*}

\begin{figure*}
    
\begin{tcolorbox}[colback=lightgray!20,colframe=darkgray!80,title=Prompt 5: Rewrite queries]
\small
system:

You are an expert Query Rewriting Agent for a Tool Retrieval System. Your goal is to decompose a user request into a structured multi-field retrieval query that aligns with our tool documentation schema.

\# Target Tool Schema

The tools you are retrieving are structured as follows:

- **description**: Formal technical summary of the tool's function.

- **parameters**: Definitions of required inputs (names, types, roles).

- **response**: Description of the data returned (entities, status, structure).

- **examples**: List of `user\_intent` strings showing how real users phrase the request.

\# Input Context

You will be provided with:

1. **User Query**: The raw user request.

2. **Partial Tool Descriptions in the Toolset**: A list of potentially relevant tool descriptions retrieved via keyword search. Use these to align your vocabulary with the available toolset.

3. **Tool View**: A human-friendly flattened view of the tool. This is provided to help you better understand the tool schema.

\# Your Output Task

Generate a JSON object with exactly two top-level fields:

1. **`tool\_needs`** (List[Dict]): 

 \ \ \ \ - Decompose the request into one or more sequential actions. Each dictionary must contain:
 
 \ \ \ \ - `user\_intent` (str): A natural language phrasing of this step, optimized to match the **examples** field.
 
 \ \ \ \ - `tool\_description` (str): A formal, technical description of the action, optimized to match the **description** field.
 
 \ \ \ \ - `expected\_response` (str): An abstract description of the expected data or outcome, optimized to match the **response** field. If no data is returned, write "Don't need response.".

2. **`extracted\_arguments`** (Dict[str, str]):

 \ \ \ \ - A single dictionary containing all parameters extracted from the query.
 
 \ \ \ \ - **Key**: A logical parameter name.
 
 \ \ \ \ - **Value**: An **abstract description** of the parameter's semantic role and data type.
 
 \ \ \ \ - **Constraint**: NEVER include concrete values from the query (e.g., do not write "Alice" or "123"). Instead, write "The target username" or "Numeric ID".

\# Response Format

Return ONLY the JSON object. Do not add markdown formatting or explanations.

\# Example

Query: "Find a flight to London and book it for John Doe."

Output:

\{

\ \ \ \   "tool\_needs": [
  
\ \ \ \  \ \ \ \      \{
      
\ \ \ \ \ \ \ \ \ \ \ \            "user\_intent": "I want to search for available flights to a specific city.",
          
\ \ \ \ \ \ \ \  \ \ \ \          "tool\_description": "Retrieve a list of flight itineraries based on destination and passenger criteria.",
          
 \ \ \ \ \ \ \ \ \ \ \ \          "expected\_response": "A collection of flight objects containing flight IDs, prices, and schedules."
          
\ \ \ \  \ \ \ \      \},
      
 \ \ \ \ \ \ \ \      \{
      
 \ \ \ \ \ \ \ \  \ \ \ \         "user\_intent": "Book a ticket for the selected flight using passenger details.",
          
 \ \ \ \ \ \ \ \  \ \ \ \         "tool\_description": "Book a ticket for the selected flight using passenger details.",
          
 \ \ \ \ \ \ \ \  \ \ \ \         "expected\_response": "A confirmation object including a booking reference number and status."
          
 \ \ \ \  \ \ \ \     \}
      
 \ \ \ \  ],
  
 \ \ \ \  "extracted\_arguments": \{
  
  \ \ \ \ \ \ \ \     "destination": "The name of the arrival city or airport code.",
      
  \ \ \ \ \ \ \ \     "passenger\_name": "The full name of the person traveling.",
      
 \ \ \ \ \ \ \ \      "flight\_id": "A unique identifier for the selected flight itinerary."
      
 \ \ \ \  \}
  
\}

\vspace{3mm}

user:

\# Partial Tools in the Toolset

The following tool descriptions were roughly matched via keyword search. 

\{retrieved\_tool\_descriptions\}

\# Tool View

It's important to note that this is just a tool example; it doesn't mean you need to generate tools exactly as shown.

\{tool\_view\}

\# User Query

\{user\_query\}

\# Task

Return **only** the JSON object. Do not include markdown formatting or explanations.
\end{tcolorbox}

\end{figure*}

\end{document}